\documentclass{aa}
\usepackage{graphicx,amsmath}
\usepackage{txfonts}

\usepackage{natbib} 
\bibpunct{(}{)}{;}{a}{}{,}

\begin{document}

 \title{The signature of chromospheric heating in Ca II H spectra}

   \author{C. Beck\inst{1,2} \and W.Schmidt\inst{2} \and R.Rezaei\inst{2} \and W.Rammacher\inst{2}}
        
   \titlerunning{Signature of chromospheric heating in Ca II H spectra}
  \authorrunning{C. Beck et al.}  
\offprints{C. Beck}

   \institute{Instituto de Astrof\'{\i}sica de Canarias
     \\
        \and Kiepenheuer-Institut f\"ur Sonnenphysik\\
     }
 
\date{Received xxx; accepted xxx}

\abstract{The heating process that balances the solar chromospheric energy
  losses has not yet been determined. Conflicting views exist on the source of the
  energy and the influence of photospheric magnetic fields on chromospheric heating.}{We analyze a 1-hour time series of cospatial Ca II H
  intensity spectra and photospheric polarimetric spectra around 630\thinspace
  nm to derive the signature of the chromospheric heating process in the
  spectra and to investigate its relation to photospheric magnetic fields. The data were taken in a quiet Sun area on disc center without strong magnetic activity.}{We have derived several characteristic quantities of Ca II H to define
  the chromospheric atmosphere properties. We study the power of the Fourier transform at different wavelengths and the phase relations between them. We perform  local thermodynamic equilibrium (LTE) inversions of the
  spectropolarimetric data to obtain the photospheric magnetic field, once
  including the Ca intensity spectra.}{We find that the emission in the Ca II H line core at locations without detectable photospheric polarization signal is due to waves that propagate in
  around 100 sec from low forming continuum layers in the line wing up to the
  line core. The phase differences of intensity oscillations at different
  wavelengths indicate standing waves for $\nu <$ 2 mHz and propagating waves
  for higher frequencies. The waves steepen into shocks in the
  chromosphere. On average, shocks are both preceded and followed
  by intensity reductions. In field-free regions, the profiles show emission
  about half of the time. The correlation between wavelengths and the decorrelation time is significantly higher in the presence of magnetic fields than for field-free areas. The average Ca II H profile in the presence of magnetic fields contains emission features symmetric to the line core and an asymmetric contribution, where mainly the blue H2V emission peak is increased (shock signature).}{We find that acoustic waves steepening into shocks are responsible for the emission in the Ca II H line core for locations without photospheric magnetic fields. We suggest using wavelengths in the line wing of Ca II H, where LTE still applies, to compare theoretical heating models with observations.}
\keywords{Sun: chromosphere, Sun: oscillations}
\maketitle
\section{Introduction}
The discovery of the flash spectrum of the sun in the late 19th century led astronomers to call this colorful layer of the solar atmosphere the \textit{chromosphere}. Since the emission lines in the chromospheric spectra have 
to form in a  hotter medium than the visible photosphere, the chromospheric temperature stratification and its heating process became the first challenge to our understanding of the outer solar atmosphere. The temperature rise in the chromosphere is a direct consequence of the rare (radiative) interactions in a low-density medium, where departure from LTE is significant. Including non-LTE is a key ingredient in all models of the outer solar atmosphere \citep[][and references therein]{fontenla_etal_06}.

The dominant heating mechanism in the chromosphere has been a matter of discussion in the last 60 
years~\citep{nar_ulm_96}. \cite{biermann_48} was one of the first to suggest that mechanical heating prevents the chromosphere from rapidly cooling down to below the photospheric temperature. 
Although the presence of waves in the solar atmosphere is well established, their importance for the chromospheric energy balance is under debate. \citet{rammacher+ulmschneider1992}, for instance, modeled wave propagation in the solar atmosphere using a 1-D code with various initial velocity fields. They obtained 3-min like oscillations using a short-period driver with 40 seconds cadence by the process of shock overtaking or shock merging. Their results depended slightly on the exact shape and periodicity of the initial wave field, 
supposedly present from the interference of the acoustic waves  permeating the 
solar photosphere. Contrary to that, \citet{fossum+carlsson2005,fossum+carlsson2006} have recently suggested that 
the acoustic wave power is not sufficient to supply chromospheric energy losses \citep[see also][]{wedemeyer+etal2007a}.

The role of magnetic fields for the chromospheric energy balance is also unclear. The strong flux concentrations at the boundaries of (super)granules can be easily identified through their emission in chromospheric lines (``{\it chromospheric network}''), but the importance of weaker magnetic fields for the chromosphere is unknown, as well as why there is permanent enhanced emission in the network. \citet{kalkofen1996} suggest collisions between flux concentrations and granules as the initial driver of the oscillations, which would only be an indirect influence. 
\citet{reza+etal2007} decomposed average Ca profiles into a non-heated, a non-magnetically, and a magnetically heated component. They conclude that the magnetic heating depends on the photospheric magnetic flux to some extent, but in total is weaker than the non-magnetically heating. 
\begin{figure*}
\resizebox{17cm}{!}{\includegraphics{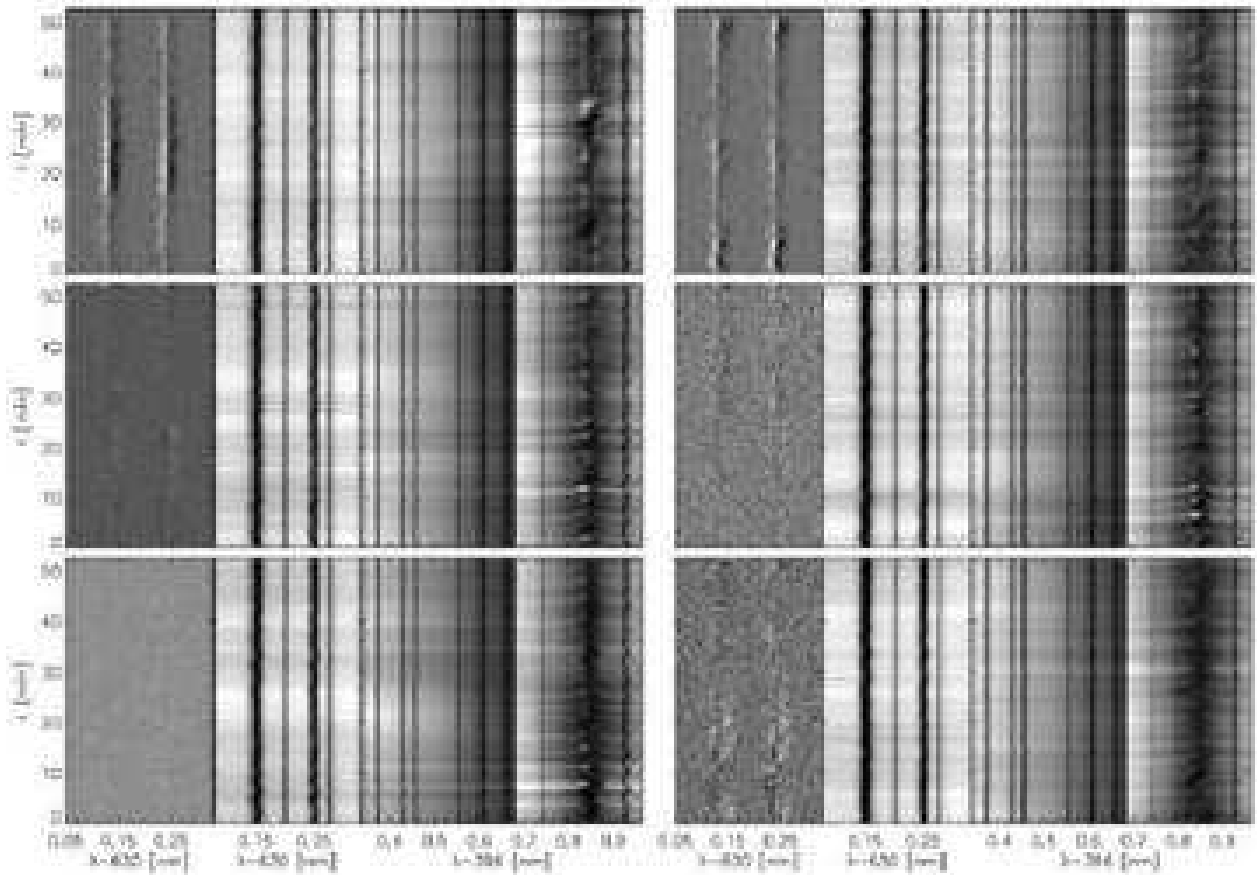}}
\caption{Six examples of the temporal evolution at fixed positions along the slit. Time is increasing from bottom to top for each example. {\em Right to left in each image}: Ca spectra; intensity and Stokes V spectra at 630\thinspace nm. The Ca spectra are displayed in two separately scaled parts, to enhance the visibility of the line core.\label{fig2}}
\end{figure*}

Few strong spectral lines are suited to observe the chromosphere from the ground (Ca II H and K, the Ca II triplet around 850\thinspace nm, H$\alpha$, Helium 1083\thinspace nm). 
Most of these lines are very broad and deep, often observed using broad-band filters with very limited spectral resolution. From space, additional low-forming emission lines and continuum windows are accessible \citep[e.g., in the UV at 1600 nm;][]{fossum+carlsson2006,dewijn+etal2007}. 
Hence, the analysis of the chromosphere is based 
on either high-resolution spectra with low temporal cadence and a limited field-of-view (FOV), or on two-dimensional data with large FOVs and poor spectral resolution. Only recently, 2-D observations with high spectral resolution have been carried out on spectral lines in the chromosphere and the upper photosphere \citep[][using IBIS]{vecchio+etal2007}. 

Several approaches have been taken in simulations to examine the mechanical heating and the shock propagation in the upper solar atmosphere: 
1-D dynamical simulations of waves in the solar atmosphere \citep[e.g.][]{rammacher+ulmschneider1992}, and 2-D and 3-D simulations of the solar atmosphere without an artificial excitation of waves \citep[e.g.][]{wedemeyer+etal2004}. 
Whereas temporally and spatially averaged spectra can be reproduced by static temperature 
stratifications with a temperature minimum between photosphere and 
chromosphere~\citep[][and its variants]{vernazza+etal1981}, the dynamical evolution of the chromosphere leads to various physical states 
and even more observed profile shapes (due to the integration across different 
physical states along the line-of-sight in a single profile). 
All numerical simulations predict cool episodes in the chromosphere 
which is in contradiction with a full-time hot chromosphere. 
The discrepancy between spectra with high spatial and temporal resolution from observations or simulations and the classical chromospheric models has been a matter of strong debate \citep{fontenla_etal_07}. 

One of the few examples of a successful combination of observations and theory is the interpretation of the bright points or bright grains seen in the Ca II H and K lines. These grains appear as short-lived small-scale intensity enhancements in the H and K line cores all over the solar surface, in internetwork areas seemingly devoid of magnetic fields, but also on the locations of network fields at the boundaries of supergranules. The grains can repeat a few times with a cadence of around 200 sec \citep[e.g.,][]{kneer+uexkull1995}, but the process is stochastic. \citet{rutten+uitenbroek1991} concluded that H$_{\rm 2V}$ grains, where the blue emission peak of the Ca II H line is increased, and the less frequent H$_{\rm 2R}$ grains are produced by the collision of p-mode oscillations, an event inside the photospheric wave field. The direct relation of the grains to the photosphere was later proven by \citet{carlsson+stein1997}. They successfully reproduced the temporal evolution of observed Ca II H spectra in a dynamical 1-D simulation, employing a photospheric ``piston'' with velocities taken from a line blend in the wing of Ca II H. Their calculations have, however, generally yielded shocks that are too strong and temperatures that are too extreme to seem realistic for the Sun, but this is presumably due to the 1-D ansatz.

In this contribution, we used data from the POLIS instrument \citep{beck+etal2005b} to study the signature of the chromospheric heating process in Ca II H spectra. The polarimetric channel of POLIS at 630\thinspace nm allowed us to localize photospheric magnetic fields, removing one of the big unknowns when analyzing Ca intensity spectra or filtergrams. Using the cospatial and cotemporal intensity spectra in the Ca II H line and vector-polarimetric spectra at 630\thinspace nm (Sect.~\ref{sect2}), we analyzed a 1-hr time series on both the statistical properties of the chromospheric emission and its relation to photospheric fields (Sect.~\ref{sect3}). We especially investigated the variation of properties with wavelength in the wing of the Ca II H line, which samples the layers from the photosphere to the line core in the chromosphere. We divided the FOV into three subfields: emission with and without strong photospheric magnetic field and very quiet regions. We compared the average profiles of the three regions, assuming them to contain different heating contributions (no heating, purely acoustic, acoustic and magnetic heating) in Sect.~\ref{sect4}. We then studied the evolution of individual and spatially averaged profiles connected to shock events (Sect.~\ref{sect5}). The findings are summarized and discussed in Sect.~\ref{sect6}.

\begin{figure}
\resizebox{8cm}{!}{\includegraphics{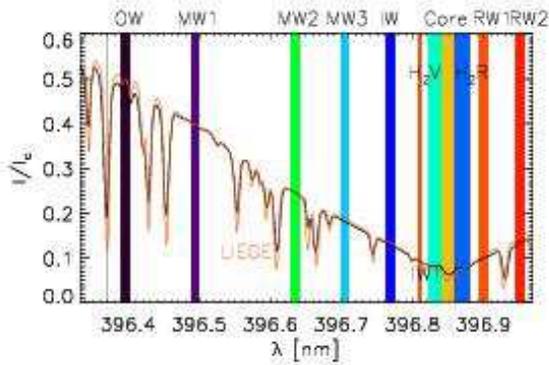}}
\caption{Average Ca spectrum with the wavelength bands for the derived
intensities. {\em Black line}: average Ca spectrum. {\em Orange line}: LIEGE atlas reference profile. The thin vertical line marks the spectral line at 396.38\thinspace nm. \label{fig3}}
\end{figure}
\begin{table}
\caption{The wavelength bands defined in the line wing and line core.\label{tab1}}
\begin{tabular}{lcc}\hline\hline
name & shortcut& wavelength $\pm$ width [nm]\cr\hline
outer wing & OW & 396.397 $\pm$ 0.005 \cr
middle wing 1 & MW1 & 396.494 $\pm$ 0.005\cr
middle wing 2 & MW2 & 396.634  $\pm$ 0.005 \cr
middle wing 3 & MW3 & 396.702  $\pm$ 0.005\cr
inner wing  & IW & 396.766  $\pm$ 0.005\cr
inner wing & IW1 & 396.807 $\pm$ 0.002 \cr
H$_{\rm 2V}$ & & 396.832  $\pm$ 0.012  \cr
Core & & 396.847 $\pm$ 0.007 \cr
H$_{\rm 2R}$ & &396.856 $\pm$ 0.012   \cr
red wing 1 & RW1 & 396.892 $\pm$ 0.005\cr
red wing 2  & RW2 & 396.950 $\pm$ 0.005\cr
H-index & & 396.848 $\pm$ 0.050\cr\hline
\end{tabular}
\end{table}
\section{Observations, data reduction, and data alignment\label{sect2}}
We observed a quiet Sun area on disc center on 24 July 2006,
from UT 08:11:50 until 09:06:44, with the POlarimetric LIttrow Spectrograph
\citep[POLIS,][]{beck+etal2005b} at the German Vacuum Tower Telescope on Tenerife. There was no sign of significant magnetic activity near disc center on that day. The full data set consisted of a scan of 4
steps with 0\farcs5 step width that was repeated more than 150
times\footnote{Overview on the full data set can be found at
  http://www.kis.uni-freiburg.de/$\sim$cbeck/POLIS\_archive/.}. The slit width
corresponded to 0\farcs5, and the integration time per scan step was around 5
seconds. The Kiepenheuer-Institute adaptive optics system
\citep{vdluehe+etal2003}, which was used to improve image quality, lost
tracking after around 1 hour. For the present study, we thus selected only the first 150 repetitions of the scan for analysis.
\begin{figure}
\centerline{\resizebox{6.5cm}{!}{\includegraphics{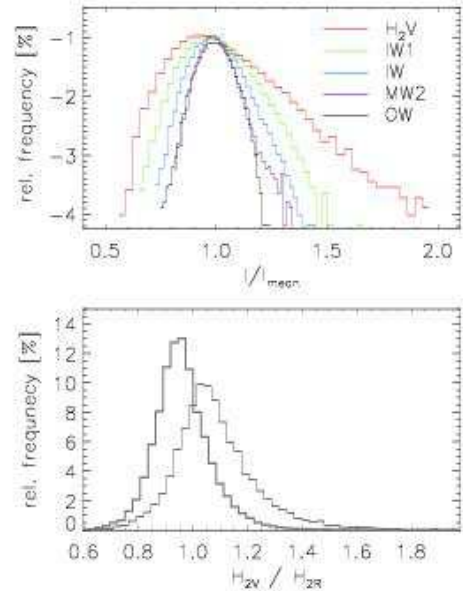}}}
\caption{{\em Top}: Intensity histograms for some of the wavelength bands. The
  intensity map in each wavelength has been normalized to its mean value
  beforehand. {\em Bottom}: Ratio H$_{\rm 2V}$/H$_{\rm 2R}$ using the wavelength bands ({\em thick grey}) or the peak intensities ({\em thin black}).\label{fig4}}
\end{figure}

The spectra of both POLIS channels ({\em blue}, Ca II H 396.8\thinspace nm, intensity profiles; {\em red}, 630\thinspace nm, Stokes $IQUV$) were reduced with the usual flatfield and polarimetric calibration procedures \citep{beck+etal2005a,beck+etal2005b}. The Ca spectra were additionally corrected for the transmission curve of the order-selecting interference filter in front of the camera. The Ca spectra were normalized afterwards to the LIEGE spectral atlas reference profile \citep{delbouille+etal1973}. For the Ca spectra, the wavelength scale of the LIEGE profile was adopted; for the red channel, the wavelength scale was set to have the line core of 630.15 (630.25)\thinspace nm in the average profile of the full data set at a convective blueshift of -180 (-240) ms$^{-1}$. The intensity increase during the observations due to the rise of the Sun was removed separately in both channels by a fit of a straight line to average intensities along the slit. 

Due to the differential refraction in the earth atmosphere \citep[e.g.][]{reardon2006}, the spectra of the two POLIS channels were not fully cospatial and cotemporal. The spatial displacement of the two wavelengths perpendicular to the slit was of the order of 2$^{\prime\prime}$, given the date of the observations, the slit orientation, and the location of the first coelostat mirror (cf.~Appendix \ref{appA}). Hence, only the first step in the blue channel was cospatial to the last of the four steps in the red channel in each scan. Discarding all steps without cospatial spectra, the data thus finally reduce to a time series of around 52 min duration with fixed slit and with around 21 seconds cadence, but in two wavelengths. The red channel was taken 21 seconds {\em later} than the blue. As no alignment in the direction perpendicular to the slit is possible (no data available), the data have been aligned only along the slit by cross-correlation of the tempo-spatial intensity maps. After alignment, a cospatial slit of 209 pixels with a sampling of 0\farcs29 per pixel remained.
\begin{figure*}
\resizebox{17.3cm}{!}{\includegraphics{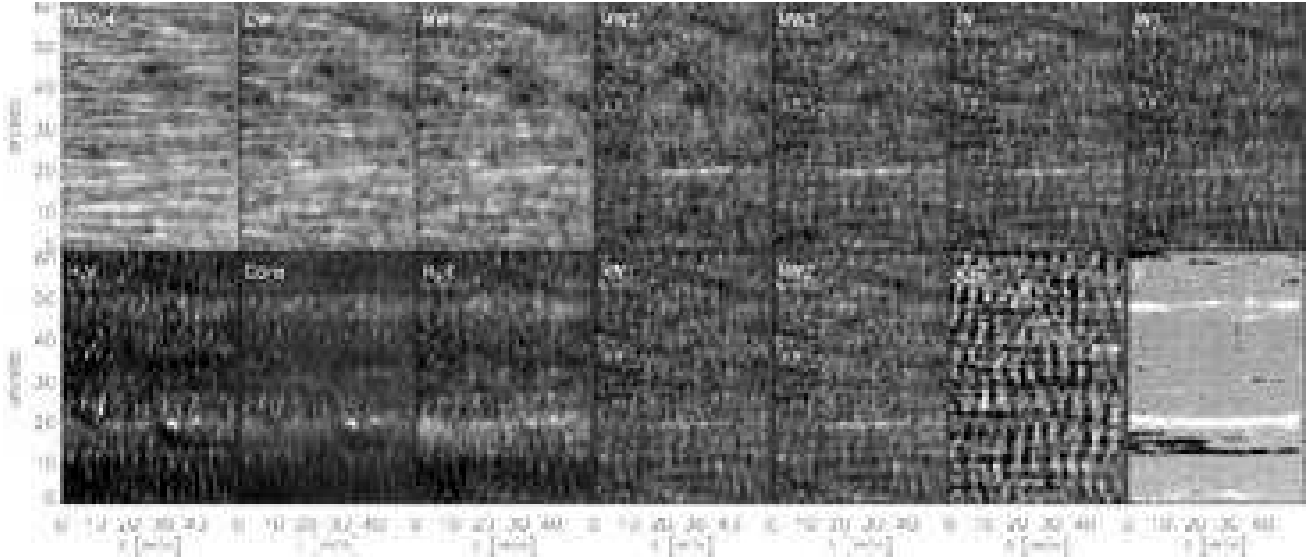}}
\caption{Spatio-temporal maps of continuum intensity in the red channel ({\em
    top left}), in the intensity bands defined in Table \ref{tab1}, LOS
  velocity of 630.25\thinspace nm (dark corresponds to blue shifts), and
  integrated signed Stokes $V$ signal ({\em bottom right}). \label{fig5}}
\end{figure*}
\section{Data analysis \& results\label{sect3}}
The data consists of the intensity profiles of the blue channel, and the Stokes vector measurements in the red channel. Figure \ref{fig2} shows six examples of the temporal evolution of spectra at different locations along the slit during the full time series. These examples serve to show the kind of informations that can be extracted from the data set: the temporal evolution of the thermodynamics at several heights in the atmosphere from the blue channel, and the photospheric magnetic and velocity fields from the polarimetric red channel.
\subsection{Ca II H spectra}
Similar to \citet{reza+etal2007}, we defined several wavelength bands in the
Ca spectra, going from outermost wing across the line core to the red
wing of the Ca spectrum (cf.~Fig.~\ref{fig3}). Table \ref{tab1} contains the
wavelength bands in detail. The bands were mainly choosen to encompass
continuum windows, besides those in the Ca line core (H$_{\rm 2V}$,
core, H$_{\rm 2R}$). Note that our definition is different from the one used
in previous literature. The regions used for H$_{\rm 2V}$ and H$_{\rm 2R}$ in
fact touch each other; the wavelength bands for the emission peaks thus include the
left and right halves of the absorption core, respectively. As both core and
emission peaks can show large spectral displacements (Fig.~\ref{fig2}), we
think they are covered better by this extended range. We also used the
H-index as an emulated 1-{\AA}-filter centered on the line core. In some methods employed to obtain information, e.g.~Fourier power or correlation matrices, each wavelength of the Ca spectrum was used individually. For all Ca spectra with two reversals in the intensity profile, we determined the amplitude and location of the H$_{\rm 2V}$ and H$_{\rm 2R}$ emission peaks. We derived the line-core velocity of Fe I at 396.38\thinspace nm as a measure of the photospheric flow field. We chose this line because it is isolated, deep, and far separated from the Ca II H line core. As the line is contained in the Ca line wing, it is perfectly cospatial and cotemporal to the rest of the Ca spectrum. 

The upper panel of Fig.~\ref{fig4} shows the intensity histograms in some of
the wavelength bands. Two trends can be seen going from the wing towards the
line core: the histograms become broader, and more and more asymmetric with an
extended tail of high intensities. \citet{lennarts+wedemeyer2005} obtained the intensity distribution at a wavelength of 396.74\thinspace nm from simulations with the Co$^5$bold code, corresponding to the IW in the present paper. Comparing to their Fig.~2, we caution that the effects of degrading the (simulations') spatial resolution is similar to moving in wavelength from, e.g,  H$_{\rm 2V}$ to IW1. Intensity histograms of filtergram observations then will be very sensitive to both the spatial resolution and the exact location and width of the filter used.

\citet[][R05]{rammacher2005} suggested using the ratio of H$_{\rm
  2V}$/H$_{\rm 2R}$ to determine the shape and amplitude of the acoustic power
spectrum presumably heating the chromosphere. The lower panel of
Fig.~\ref{fig4} shows the histograms of the ratio using the wavelength bands,
or all locations, where a double reversal with two clear emission peaks was observed in the spectra. The two distributions are similar, centered around one, with an extended tail to high H$_{\rm 2V}$/H$_{\rm 2R}$ ratios. Compared to R05, the distributions are much smoother without isolated peaks, and smaller maximal ratios (below 2). This could be due to the difference between the 1-D calculations employed by R05, which tend to generate strong shocks, and the 3-D solar atmosphere. 
\begin{figure*}
\sidecaption
\resizebox{11cm}{!}{\includegraphics{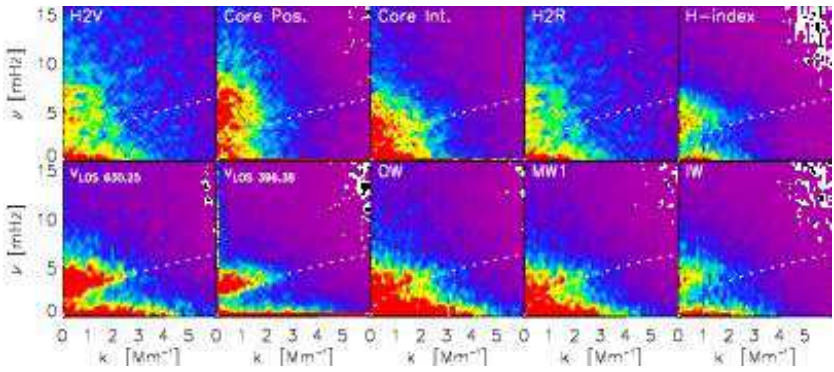}}
\caption{k-$\nu$-diagrams. {\em Bottom row, left to right}: line-of-sight
  velocity of 630.25\thinspace nm, same for 396.38\thinspace nm, outer,
  middle, inner wing. {\em Top row}: H$_{\rm 2V}$, position of Ca core
  derived from the location of minimum intensity, intensity of minimum, H$_{\rm 2R}$, H-index. The dashed line gives the fundamental mode of the p-mode oscillations ($\nu = \sqrt{g\cdot k}/2\pi$, with the solar surface gravity $g = 273$ ms$^{-2}$).\label{fig7}}
\end{figure*}
\subsection{630\thinspace nm spectra}
The spectropolarimetric spectra of the red channel were inverted with the SIR
code \citep{cobo+toroiniesta1992,cobo1998}. The inversion scheme employed a single field-free
component and straylight for profiles without clear polarization signal;
otherwise, a two-component model of one magnetic atmosphere, one field-free
component, and straylight were adopted. All quantities besides temperature were
assumed to be constant with depth. As Fig.~\ref{fig5} shows, actually only few
locations along the slit showed stronger polarization signals. This inversion
used only the 630\thinspace nm spectra, and was performed in the full field of
view.

The inversion was repeated for all time steps of 30 pixels along the slit
(y$\sim 20^{\prime\prime}$ to 30$^{\prime\prime}$ in Fig.~\ref{fig5})
including the Ca spectra, with identical settings to those before. The aim was
to investigate whether an LTE inversion can still capture the propagation of waves
and shocks in the lower atmospheric layers. Appendix \ref{appb} shows some
examples of the observed and best-fit profiles of this inversion setup. From
the satisfactory reproduction of the observed spectra -- excluding the actual
Ca line core and maybe the IW1 band -- we conclude that the continuum
bands defined should be accessible by an LTE inversion \citep[cf.][for the Ca II K wing]{owocki+etal1980}. However, we refrain from using these inversion results at present before a rigid investigation of their reliability.  The line-core velocities of the Fe I lines at 630.15\thinspace nm and
630.25\thinspace nm were determined as additional measurements of the
photospheric velocity field. The velocity dispersion of the red channel of 0.7
kms$^{-1}$ per pixel is about two times smaller than in the blue channel,
giving a better velocity resolution.
\begin{figure}
\centerline{\resizebox{8.8cm}{!}{\includegraphics{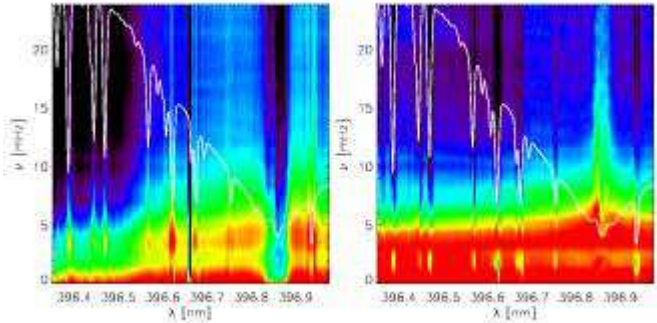}}}
\caption{Power as function of wavelength and frequency, averaged along the slit. {\em Left}: normalized with P($\lambda, \nu$=0.3 mHz). {\em Right}: normalized with the power at 3 mHz, P($\lambda,\nu$=3 mHz).\label{fig8}}
\end{figure}
\subsection{Morphology of observed FOV}
Figure \ref{fig5} shows the temporal evolution of the intensity along the slit
in the wavelength bands of Table \ref{tab1}. Passing from continuum wavelengths to the Ca line
core, the structures visible change drastically. The tempo-spatial maps from
continuum to MW1 are dominated by the structure and the temporal evolution of
the granulation, leading to a mainly horizontally ($\equiv$temporal axis)
oriented pattern of bright and dark stripes. For all other maps, the granulation signature is completely lost and exchanged by a vertically  ($\equiv$spatial axis) oriented pattern of isolated (repeated) brightenings. The brightenings in H$_{\rm 2V}$ or core last only shortly (20-60 sec) and extend over 2$^{\prime\prime}$ to 3$^{\prime\prime}$ along the slit. Many of the brightenings on locations without strong magnetic fields are repetitive with periods of 150 sec to 250 sec.

Four magnetic elements were intersected by the slit (y$\sim 12^{\prime\prime}$,
15$^{\prime\prime}$, 20$^{\prime\prime}$, 50$^{\prime\prime}$), which are seen
all throughout the time series at approximately the same locations. Almost all
locations that were inverted with a magnetic atmosphere belong to these four
patches;  they outline network fields ($B \sim 1.3$ kG,
Fig.~\ref{fig6}). Other locations only show transient weak polarization signals.
Comparing the map of polarization signal (bottom right of Fig.~\ref{fig5}) and that of, e.g., H$_{\rm 2V}$ (bottom left), one
can discriminate between three different types of locations in the FOV. Cospatial to strong photospheric fields, one finds a quasi-static intensity increase with less signatures of oscillations, and a small halo with higher intensity on the neighboring pixels. Contrary to the field-free locations, the oscillations on the field concentrations only modulate the emission, but do not lead to its disappearance, especially in the H$_{\rm 2R}$ map. Close to fields (y$\sim 22^{\prime\prime}$ to $27^{\prime\prime}$ and $\sim 37^{\prime\prime}$ to $42^{\prime\prime}$), the periodic structure of the brightenings is most prominent (``{\em caterpillar tracks}''). In very quiet locations ($\sim 27^{\prime\prime}$ to $37^{\prime\prime}$), fewer and often
non-repetitive brightenings can be found (t$\sim 20$ min, y$\sim 32^{\prime\prime}$).
\subsection{Fourier analysis of Ca II H spectra}
To quantify the properties of the intensity and velocity oscillations, we took the Fourier transform of the tempo-spatial maps. We try to isolate the general properties of the oscillations in the full FOV, and later investigate differences between locations with or without field, or in the very quiet area, by using spatially resolved information.
\begin{figure}
\centerline{\resizebox{8cm}{!}{\includegraphics{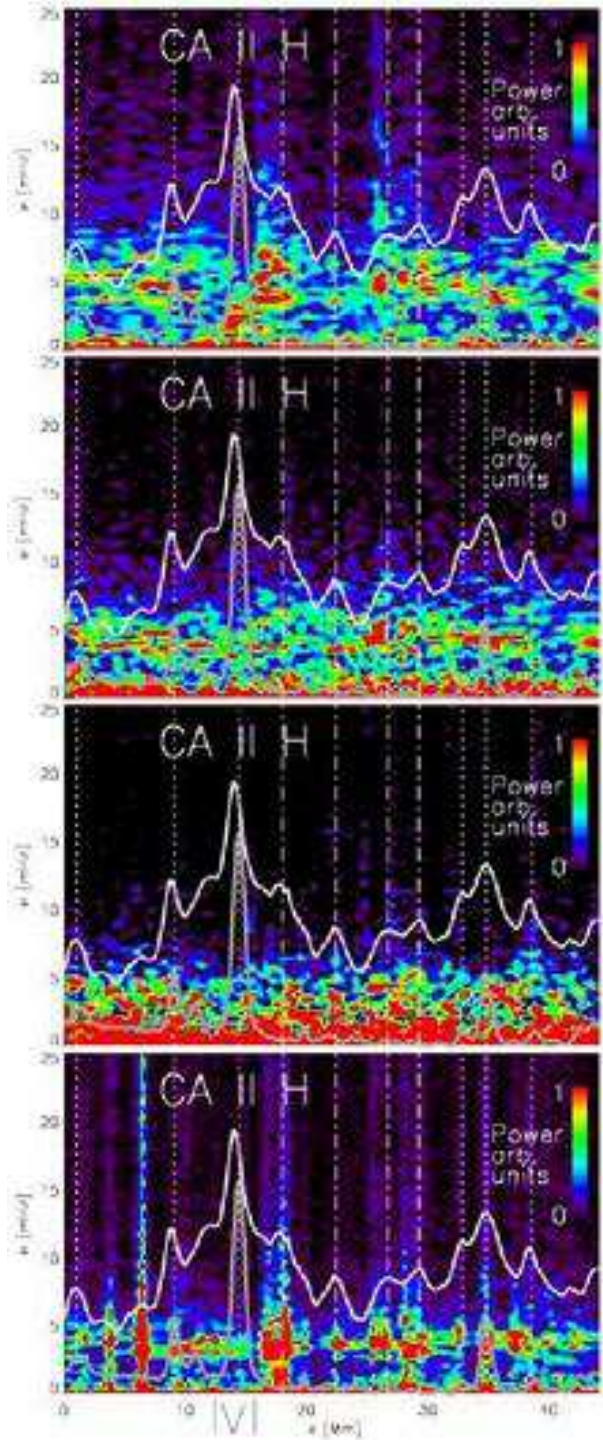}}}
\caption{Combination of spatially resolved Fourier power with curves of average H-index ({\em white}) and average unsigned Stokes $V$ ({\em grey}) along the slit. {\em Top to bottom}: power in H$_{2V}$, IW,OW, $v_{\rm LOS}$ 630.25\thinspace nm. {\em Dotted lines} mark enhanced polarization signal, {\em dash-dotted lines} enhanced Ca II H emission without strong polarization signal.\label{fig9} }
\end{figure}
\paragraph{k-$\nu$-diagrams} Power spectra of solar oscillations as function
of spatial and temporal frequency are a well known tool used in many
studies.  For the chromospheric Ca II H line few examples using spectroscopy have been published since \citet{cram1978}; in most cases, filtergrams were used \citep[][or other publications on data from the Dutch Open Telescope]{rutten+etal2004a}. k-$\nu$-diagrams have often been presented for the Ca II K line \citep{kneer+uexkull1983,dame+etal1984,steffens+etal1995} or H$\alpha$ \citep{kneer+uexkull1985}. \citet{lennarts+wedemeyer2005} show the only k-$\nu$-diagram from a simulation, for a wavelength of 396.74\thinspace nm. We will concentrate here mainly on the changes with wavelength in the shape of the k-$\nu$-diagrams. The velocities and intensities up to MW1 show strongest power in the 3-4 mHz range, corresponding to the photospheric 5 min oscillations (Fig.~\ref{fig7}). For the inner wing, the frequencies with high power start to increase. For all quantities from the Ca line core, significant power can be found up to frequencies of 10 mHz. The power in the quantities from the line core of calcium is usually spread over a broad frequency range. The power distribution is similar to the one found by \citet[][Figs.~5.11 and 5.12, p.~84ff]{woeger2006}, who used short-exposed narrow-band Ca II K core images. Only for the broad-band H-index is there a prominent peak near 3 min. This could indicate that, with the increased spatial and spectral resolution and higher S/N ratio in recent observations, more of the fine-structure of the dynamic chromospheric evolution is detected, whereas previous observations using broad-band filtergrams have mainly traced the bright grains with their characteristic 3-min repetition time. Only OW and MW1 do not show reduced power from 1 to 2 mHz, which all other graphs exhibit. The k-$\nu$-diagrams of the line-core velocity oscillations of the Fe I lines (lower left) match more closely to the that of IW1 than that of the outer wing, reflecting the formation of the line cores above the continuum.
\begin{figure}
\resizebox{8cm}{!}{\includegraphics{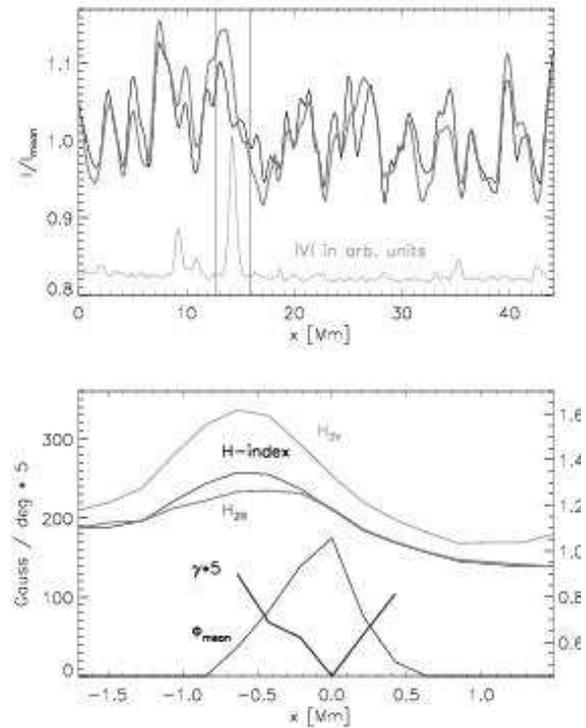}}
\caption{{\em Top}: continuum intensity along the slit at 630.4\thinspace nm (red channel, {\em black}) and in the wing of calcium (blue channel, {\em grey}). The Stokes V signal along the slit is overplotted, the strongest concentration is marked by two vertical lines. {\em Bottom}: blow-up of the field concentration. Magnetic flux, $\Phi_{\rm mean}$, and field inclination, $\gamma$, are shown at the bottom, intensities in H$_{2V}$, H-index, and H$_{\rm 2R}$ at the top.\label{fig10} }
\end{figure}
\paragraph{Average power spectra vs.~wavelength} To study the dependence of power on wavelength, $\lambda$, without regard to the spatial extent, we also calculated the k-$\nu$-diagrams for all wavelength points in the Ca spectrum, and averaged over the spatial frequencies. Figure \ref{fig8} displays the power as function of $\lambda$ for two different normalizations of the Fourier power: using the power at the first non-zero frequency, P($\lambda, \nu$=0.3 mHz), or at 3 mHz, P($\lambda,\nu$=3 mHz). For the first normalization, the power in the Ca line core is strongly reduced compared with the wing; in the second case, the increased power at higher frequencies near the Ca line core shows up prominently. All line cores of photospheric spectral lines also show more medium-frequency power (3-4 mHz$\equiv$ 4-5 min) than their close-by continuum.
\begin{figure}
\resizebox{6cm}{!}{\includegraphics{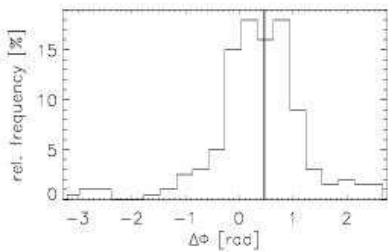}}
\caption{Histogram of phase differences between H$_{\rm 2V}$ and IW1 at $\nu$ = 4.8 mHz. The vertical line marks the center of gravity of the distribution.\label{fig11}}
\end{figure}
\begin{figure*}
\sidecaption
\resizebox{11.5cm}{!}{\includegraphics{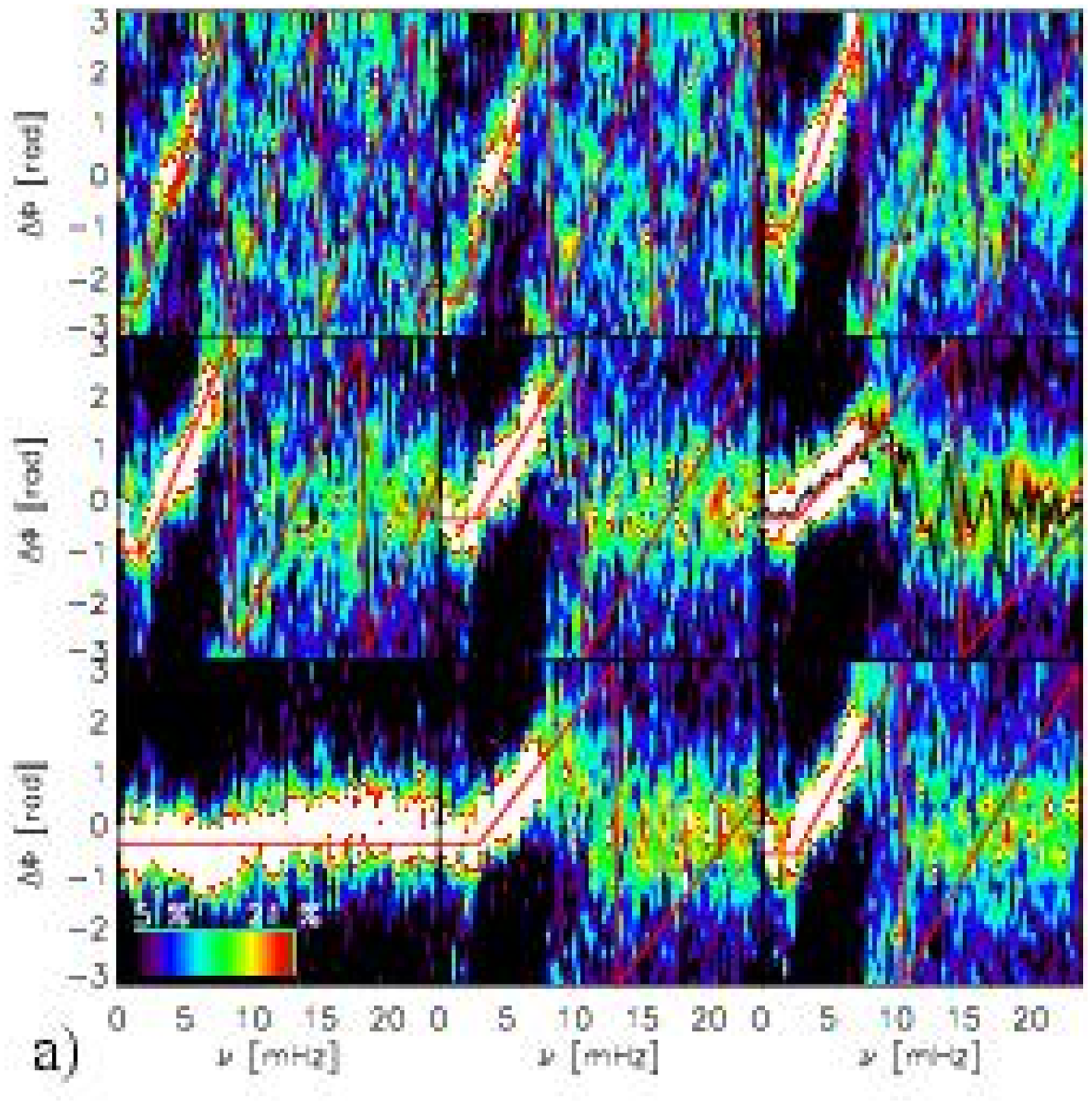}}
\caption{(a) Fourier phase differences between the H$_{\rm 2V}$ emission peak and different wavelengths. For each frequency, the color code (cf.~lower left
  corner) shows the relative occurrence of a given phase shift. {\em Red lines}
  are a manual fit of piecewise defined straight lines to the data. Three
  parameters were used to construct the curves: a variable offset for $\nu$ =
  0 mHz, the minimum frequency of propagating waves ($\sim$ 1-2 mHz), and the
  slope of the dispersion relation. {\em Top row, left to right}: OW,
  MW1,MW2. {\em Middle row}: MW3, IW, IW1. {\em Bottom row}: core intensity,
  RW1, RW2. For IW1, the center of gravity of the phase difference
  distribution is overplotted in black. (b) Phase difference between H$_{\rm 2V}$ and $v_{\rm LOS}$ 630.15\thinspace
  nm. The phase differences have not been corrected for the 21 sec time
  difference between the blue and red channel that introduces an additional
  frequency-dependent phase shift.\label{fig12}}\vspace*{-11.6cm}\mbox{\hspace*{13cm}\resizebox{4.5cm}{!}{\includegraphics{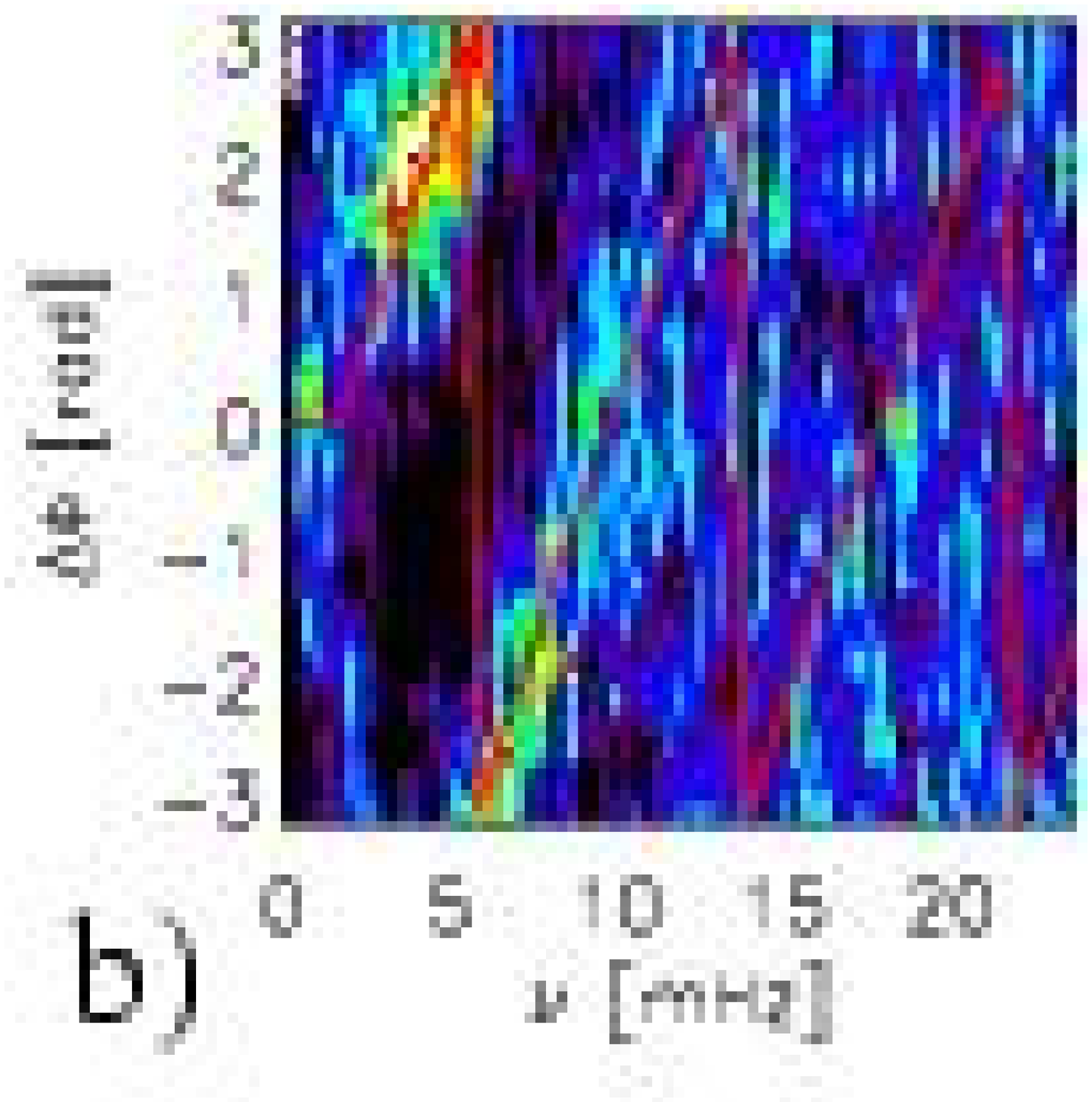}}  }\vspace*{7.cm}
\end{figure*}
\paragraph{Spatially resolved power spectra} To investigate the influence of
the photospheric fields on the power spectrum, we calculated the power
spectrum of each pixel along the slit separately \citep[cf.][their Figs.~4 and 5]{lites+etal1993}. Figure \ref{fig9} shows the
spatially resolved power spectrum of the H$_{2V}$ peak, inner and outer wing, and of the line-of-sight velocity of Fe I at 630.25\thinspace nm. For
comparison, the unsigned integrated Stokes $V$ signal, $\int |V(\lambda)| d\lambda$ (grey, lower curve) and the H-index
(white, upper curve) are overplotted, both averaged over the time series. The enhanced power at high frequencies ($>$ 5 mHz) in the H$_{2V}$ peak avoids the
magnetic fields and is displaced from them by 1 to 3 Mm. This happens
at all strong flux patches (dotted lines), but especially at the strongest
one at x = 15 Mm. At the location of this field concentration, chromospheric
low-frequency power is enhanced. \citet{lites+etal1993} found the same
reduction of high-frequency power and increase of low-frequency power for
network fields; enhanced halos of 5 mHz power around network were also found
by several authors, for example \citet{krijger+etal2001}. A similar spatial displacement between magnetic fields and high-frequency power can be seen in the inner wing power spectra.

In the outer wing, the power spectrum shows very little spatial dependence at
all. The locations of the fields are not significantly different from their
surroundings. In contrast, the power in the line-of-sight velocity of the
photospheric Fe I line at 630.25\thinspace nm shows a strong spatial
variation, with several localized sources with high power at all frequencies
up to around 6 mHz (x= 6.5,8, 18,35 Mm). Only the last source coincides with strong magnetic fields. We remark that this power spectrum comes from the red channel of POLIS. Enhanced chromospheric intensity  is cospatial with either magnetic fields, or increased medium-frequency (2-6 mHz) power. Especially in the power of the photospheric velocity, each local maximum of the H-index not related to magnetic fields is correlated one-to-one with larger than average oscillation power in the photosphere (x=18,22.5,26,28 Mm).

Figure \ref{fig9} also indicates that the maximum H-index is not cospatial to
the field concentration at x=15 Mm. As the curves in Fig.~\ref{fig9} were
created from the temporal averages, we decided to use a single scan step for
closer investigation of this displacement. To
check the alignment of the red and blue channel, the upper panel of
Fig.~\ref{fig10} shows the intensity in the (pseudo-)continuum of each channel
on scan step 65. The curves clearly show that there is no systematic spatial
displacement between the channels. The lower panel shows a blow-up of the
strongest field concentration, with the values of field inclination, $\gamma$,
 and average  magnetic flux per pixel, $\Phi_{\rm mean} = B\cdot f \cdot
 \cos \gamma$, from the inversion. $f$ is the filling fraction of magnetic
 fields inside the pixel. The largest emission is displaced from the maximum
 magnetic flux by about 0.65 Mm to the left to a region, where the fields are
 more inclined ($\sim 25^\circ$). The reason for the displacement is not
 obvious, and the direction seems arbitrary at first. We note however that the
 flux concentration at x$\sim$10 Mm has the opposite polarity. Field
 lines connecting these two patches and forming a canopy between them could be the reason that the emission is displaced in that direction.
\paragraph{Phase differences} The phase differences between oscillations at different geometrical heights contain information on the presence and type of
waves. Both intensity or velocity oscillations can be related to each other \citep[e.g.][]{lites+chipman1979}. As the determination of chromospheric velocities from, e.g., the location of the Ca II H line-core position, is rather unreliable\footnote{The position of the intensity minimum can be determined, but its interpretation as velocity is doubtful.}, we only considered intensity oscillations in the following. We have first calculated the phase differences between the oscillations in the H$_{\rm 2V}$ emission peak and the wavelengths bands of Table \ref{tab1}, and later between H$_{\rm 2V}$ and all wavelengths in the blue channel.

We used a method for the creation of phase difference plots similar to
that used in previous studies \citep{lites+chipman1979,kulaczewski1992,lites+etal1993,lennarts+wedemeyer2005}. Treating each of the 209 pixels along the slit and
each wavelength (band) individually, we derived the phases, and hence, the phase differences as a function of frequency from the Fourier-transform of the intensity variation with time.  As there is a 360$^\circ$-ambiguity in the phase differences, i.e.~-190$^\circ \equiv +170^\circ$, all phase differences have been projected into the range $\pm 180^\circ$. We then calculated the histograms of relative occurence of phase differences for each frequency. Figure \ref{fig11} shows an example for the phase difference between H$_{\rm 2V}$ and IW1 at $\nu$ = 4.8 mHz. The histogram of the phase differences at a given frequency is usually well centered around a single peak. This indicates that there is a preferred phase relation between the different wavelengths, i.e.~a (retarded) connection between the oscillations. 
\begin{figure}
\centerline{\resizebox{8.8cm}{!}{\includegraphics{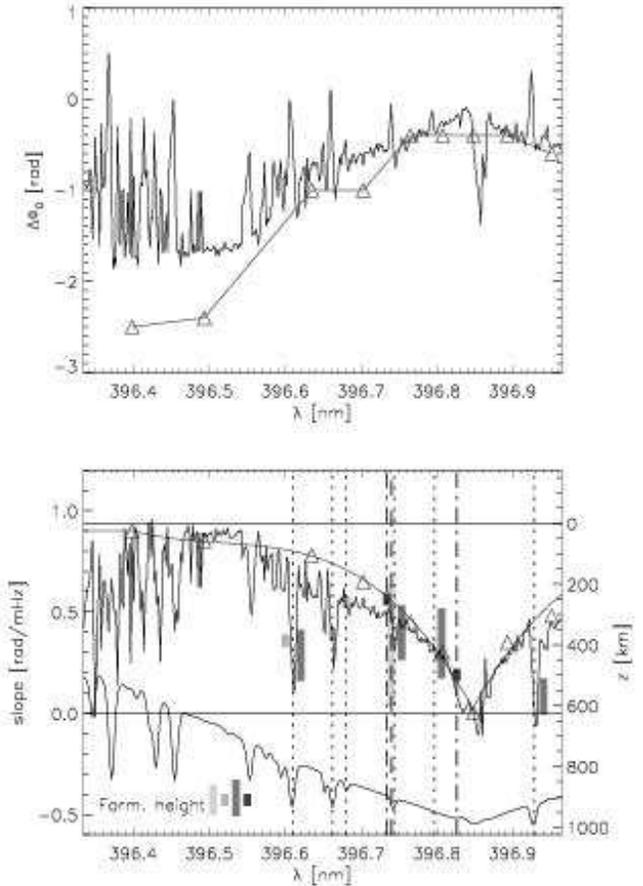}}}
\caption{Parameters of straight line fit to phase shifts. {\em Top}: phase offset for $\nu < 2$ mHz.  {\em Bottom}: slope of phase shift as function of
  frequency. {\em Solid black zig-zagging lines}: results of automatic fit. $\triangle$: results of manual fit, connected by lines for better visibility. In the bottom panel, the Ca spectrum is overplotted as
  reference. {\em Dotted vertical lines} mark spectral lines. The {\em dashed and dash-dotted vertical lines} mark ``continuum'' wavelengths (396.734~nm, 396.74~nm, 396.826~nm) outside photospheric blends. The grey shaded areas denote formation heights obtained in other studies, plotted according to the scale in km at the right. The horizontal lines mark the zero of the slope and of the geometrical height scale, respectively.\label{fig13}}
\end{figure}

Figure \ref{fig12} shows the phase differences thus obtained as function of oscillation frequency for wavelength bands going from the outer blue wing of the Ca line through the line core into the red wing. The graph contains several interesting features:
\begin{itemize}
\item Reliable phase differences can be determined from 0 up to around 10 mHz.
\item Three parameters seem to suffice for the description of the phase
  difference as a function of frequency: a phase offset for $\nu$ = 0 mHz ($\Delta\Phi_0$), a minimum frequency for propagating waves ($\nu_{\rm min}$), and the slope of the dispersion relation ($\delta(\Delta\Phi)/\delta\nu$). Using these three parameters, the observed phase differences can be reproduced by piece-wise straight lines: a constant value of $\Delta\Phi_0$ ($\nu < \nu_{\rm min}$), and a straight line with the slope $\delta(\Delta\Phi)/\delta\nu$ ($\nu > \nu_{\rm min}$).
\item The phase differences as function of wavelength separation from H$_{\rm 2V}$, going from wing to the core, then show the following trends: a reduction of $\Delta\Phi_0$ from $-\pi$ to 0, a decrease of the slope $\delta(\Delta\Phi)/\delta\nu$, and a small increase of $\nu_{\rm min}$.
\item The behavior is symmetrical around the Ca line core, i.e.~IW1 (IW) is nearly identical to RW1 (RW2). The phase shift between, e.g., IW1 and RW1 (not shown) was close to zero for all freqencies.
\end{itemize}
The phase shifts actually show more structure for low frequencies ($<$ 2mHz)
in some cases. The behaviour is similar, and can best be seen in the phase
shift between H$_{\rm 2V}$ and MW3 (Fig.~\ref{fig12}, middle row, at the
left). The phase shift starts with zero at zero frequency, decreases with
frequency until around 2 mHz, and only then starts to increase with a constant
slope. Negtive phases at low frequencies could be indicative of the presence
of gravity waves \citep[cf.][]{krijger+etal2001}.
\begin{figure}
\centerline{\resizebox{4.5cm}{!}{\includegraphics{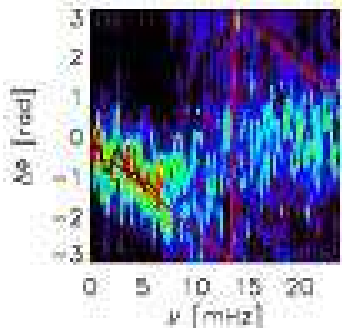}}}
\caption{Phase shift between 396.831~nm (H$_{\rm 2V}$) and 396.853~nm (line core). The straight lines resulted from the automatic fit to the COG values of the phase distributions ({\it black}).\label{fig12a}}
\end{figure}

As the plot of the phase differences between H$_{\rm 2V}$ and the nine
continuum bands showed a ``well-behaved'' relation, we tried to derive the
characteristic values ($\Delta\Phi_0,\delta(\Delta\Phi)/\delta\nu$) also for
all wavelengths $\lambda$ by an automatic method. To this extent, we calculated the phase differences between the intensity oscillations in a single wavelength pixel located in the H$_{\rm 2V}$ emission peak ($\lambda_0$ = 396.831\thinspace nm) and all other wavelengths in the spectrum of the blue channel in the same way as before. At each frequency, $\nu$, we determined the center of gravity of the phase shift histogram (cf.~Fig.~\ref{fig11}). To these values of $\Delta\Phi(\nu, \lambda-\lambda_0)$ (cf.~middle row, rightmost panel of Fig.~\ref{fig12} or Fig.~\ref{fig12a}) we fitted a straight line inside a frequency range from 1.3 mHz to around 5 mHz. This yields values of offset and slope as function of wavelength. From a visual inspection of phase differences and fit result it was seen that the offset value was less well reproduced than the slope. This is due to the fact that a fixed $\nu_{\rm min}$ was used, which affects the derived offset more strongly than the  slope. However, the results of the automatic fit were in reasonable agreement with the nine manually derived values, and yield smooth curves for both offset and slope (cf.~Fig.~\ref{fig13}). 

The curve of $\delta(\Delta\Phi)/\delta\nu$ as function of wavelength allows
for an interpretation of the smaller slope found for the continuum bands
closer to the Ca II line core: the same decrease happens for all
wavelengths located in the line cores of spectral lines in the blue or red
wing. The reason is that in the most basic formulation of propagating waves in
the solar atmosphere the phase difference is given by $\Delta\Phi(\nu)=
f(\nu) \cdot \Delta z$, with $\Delta z$ the height difference between two
layers, whose phase differences are calculated
\citep[e.g.][]{rebecca+etal2006}. $f(\nu)$ should be independent of the
wavelengths in the observations related to each other; it reflects the properties of the waves propagating in the solar atmosphere that produce the intensity variations. The change of the
slope with $\lambda$ then simply reflects the (non-linear) conversion from wavelength to geometrical height difference, $\Delta z$. 
\begin{figure}
\centerline{\resizebox{8.cm}{!}{\includegraphics{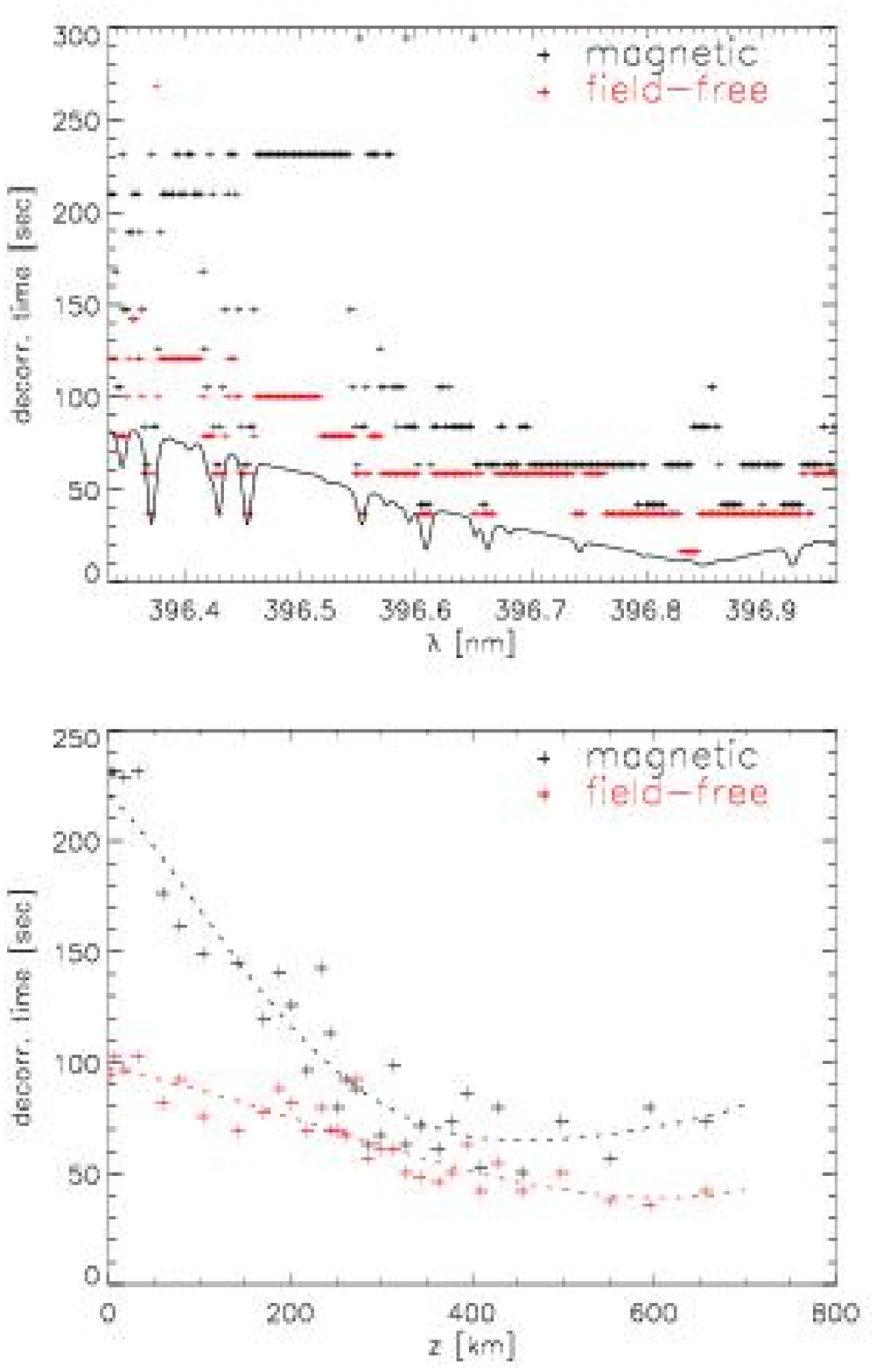}}}
\caption{{\it Upper graph}: decorrelation time as function of wavelength for magnetic (black) and field-free locations (red, slightly shifted down for better visibility). {\it Lower graph}: same as function of geometrical height. The {\it dashed} lines are a 3rd order polynomial fit for easier visualization. \label{fig16a}\label{fig16}}
\end{figure}

The relation between slope and geometrical height can then be used in the opposite direction to determine the
formation heights of spectral lines or continuum
wavelengths. \citet{deubner1974} used a similar approach to determine the
formation height of some spectral lines. We have added a second axis of
geometrical height at the right of the lower panel in Fig.~\ref{fig13}, and overplotted some formation height
ranges derived from intensity contribution functions for continuum wavelengths \citep[darkest grey,][using the FALC model]{leenaarts+etal2006}, from response functions of spectral lines \citep[H.~Schleicher, priv. comm., dark grey, published in][]{beck+etal2005b}, from phase differences \citep[grey,][]{lites+etal1993}, and from simulations \citep[light grey,][]{lennarts+wedemeyer2005}. The formation heights from these other studies are denoted by shaded areas that were drawn slightly displaced in wavelength in some cases for better visibility. We remark that the geometrical scale is determined uniquely by two references height values, z($\lambda_1$) and z($\lambda_2$), which were {\em simply chosen} in the present case. The outermost wing was set to zero height; the formation height of the line at 396.6\thinspace nm was used to yield the second value, z($396.6\thinspace {\rm nm})\sim 550$ km. With this caveat, we think the agreement between the formation heights derived from several different methods and the curve of the slope of the phase relation converted to geometrical height to be reasonable, taking into account that the derivation of height from the slope is a rather indirect method. Note that the Ca line core would also be located at only around 700 km according to the height scale. The slope actually turned to values below zero for some wavelengths in the line core (Fig.~\ref{fig12a}), opposite to the slope in the line wing. This would be required for upwards propagating waves, if the line core forms above the H$_{\rm 2V}$ peak.
\subsection{Decorrelation time}
To quantify typical time scales, we used the autocorrelation of intensity
and the decorrelation time, $\tau$, when the autocorrelation drops below 1/e \citep[e.g.,][]{lennarts+wedemeyer2005,tritschler+etal2007}. We used an automatic method to determine $\tau$ as function of wavelength for field-free and magnetic
regions. Figure \ref{fig16} shows that in the line wing, the decorrelation
time inside (outside) fields is around 200 to 250 sec (100 to 150 sec). The
decorrelation times get smaller in the chromospheric layers, and are between 21
and 63 sec near the Ca core for both field-free and magnetic regions. Note that due to the temporal sampling of 21 sec this implies a drop of the correlation from 1 to below 1/e in a single time step. Shorter decorrelation times cannot be detected with the cadence of the observations, but cannot be excluded. Directly in the Ca line core (396.86$\pm0.02$\thinspace nm), the decorrelation time reaches again around 100 seconds for the locations with magnetic fields. 
\begin{figure}
\centerline{\resizebox{8.8cm}{!}{\includegraphics{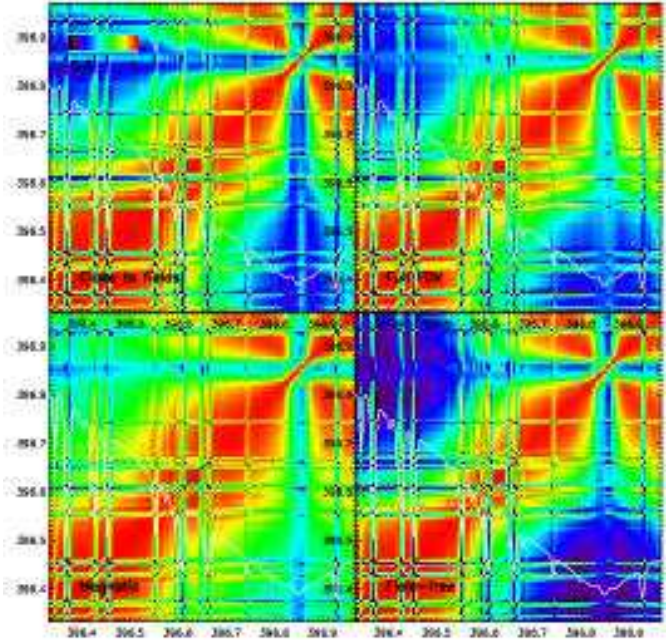}}}
\caption{Wavelength correlation matrices. {\em Clockwise, starting top
    left}: area close to magnetic fields, full FOV, field-free region,
  region with magnetic fields. The scale is $\Delta\lambda$ in nm. All
  correlations are displayed between -0.3 and 1, a color bar is given in the upper left image. \label{fig17}  }
\end{figure}
\begin{figure}
\centerline{\resizebox{8cm}{!}{\includegraphics{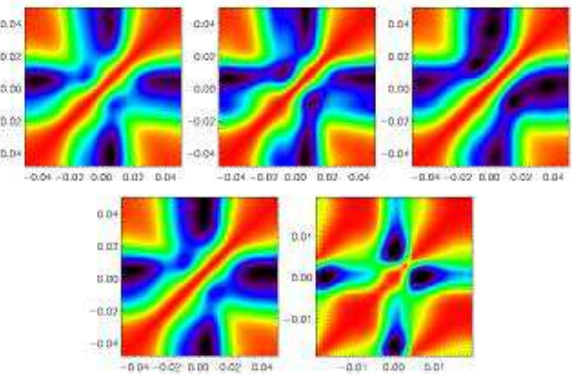}}}
\caption{Magnified view of the region around the Ca line core. ({\em Top
    row, left to right}): field-free, close to magnetic fields, magnetic. ({\em
    Bottom row, left}): full FOV, Ca line core. ({\em  Bottom row,
    right}): full FOV, Fe I at 396.45\thinspace nm.\label{fig18}}
\end{figure}

Using the relation between wavelength and height from Fig.~\ref{fig13}, we created a plot of decorrelation time as function of height (Fig.~\ref{fig16a}). With the caveat that the height scale is not very well determined, the curve may still serve for a fast comparison with simulations, as the decorrelation times of some physcial quantities like temperature, velocity, or opacity in a simulation box at a given height can be derived without spectral synthesis.
\subsection{Correlation matrices\label{corrmat}}
\citet{rammacher+etal2007} suggested investigating the amount of correlation
between different wavelengths in chromospheric spectral lines as a fingerprint
of the heating process. We thus calculated the correlation matrices for the
full wavelength range available in our spectra, and four different spatial
areas: the full field of view, field-free regions, regions with photospheric
magnetic fields, and an area close to fields, but without photospheric
polarization signal (cf.~Table \ref{tab2}). The last area has been chosen to
be next to the strongest field concentration, where the chromospheric
high-frequency power is enhanced (cf.~Fig.~\ref{fig9}). Figure \ref{fig17} shows the found 2-D
wavelength correlation matrices. The correlation is enhanced over a longer
wavelength range, if magnetic fields are present. Without magnetic fields
present, anti-correlation is found for wavelengths separated more than around
0.4\thinspace nm. The matrix for the full FOV compares well to the one given
by \citet{rammacher+etal2007} for Ca II H (their Fig.~1).
\begin{table}
\caption{Locations of the regions used in Sects.~\ref{corrmat} and \ref{sect4} in arcsec along the slit (cf.~Fig.~\ref{fig19}).\label{tab2}}
\centering
\begin{tabular}{l l l}\hline\hline
with mag.~fields & no magn.~fields & close to magn.~fields\cr\hline
 12-14, 19-22, 48-51 & 4-11, 24-43 & 22-25\cr\hline
\end{tabular}
\end{table}

The correlation matrix is highly structured around the Ca line core and
all other spectral lines. This is demonstrated in Figure \ref{fig18}, which
shows a magnified view of the Ca line core. The four different spatial
areas selected show different correlations, both in the absolute values and
the shape of the correlation matrix. Also each spectral line in the Ca
line wing produces similar patterns in the correlation matrix (lower right of
Fig.~\ref{fig18}). If the pattern is produced by the same spectrum of waves in
calcium and the other lines, this offers a good opportunity to restrain
theoretical heating mechanisms, because the blends in the line wing are better
accessible for a detailed modeling than the Ca line core itself, which requires to consider NLTE effects. 
\begin{figure}
\centerline{\resizebox{8cm}{!}{\includegraphics{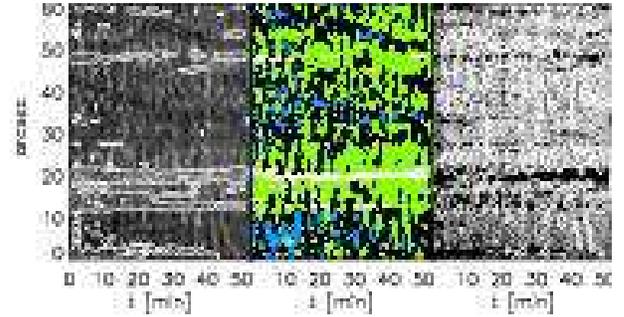}}}
\caption{{\em Left}: H-index. The upper (lower) diamond at t$\sim$10 min
  denotes the location of largest (smallest) H-index. Contours outline strong polarization signal. {\em Middle}: mask of locations with magnetic
  fields and strong emission (white), field-free emission (green), and quiet
  (blue) Ca profiles. {\em Right}: integrated Stokes $|V|$ in reversed scaling. Contours outline strong emission.\label{fig19}} 
\end{figure}
\section{Average Ca profiles: quiet, magnetic, field-free emission\label{sect4}}
In the previous sections we have concentrated on the global characteristics of
the temporal evolution of the Ca profiles. In the following we rather
investigate the shape and evolution of individual profiles, in an attempt to
identify the process leading to the emission in Ca II H.
\begin{figure}
\resizebox{8cm}{!}{\includegraphics{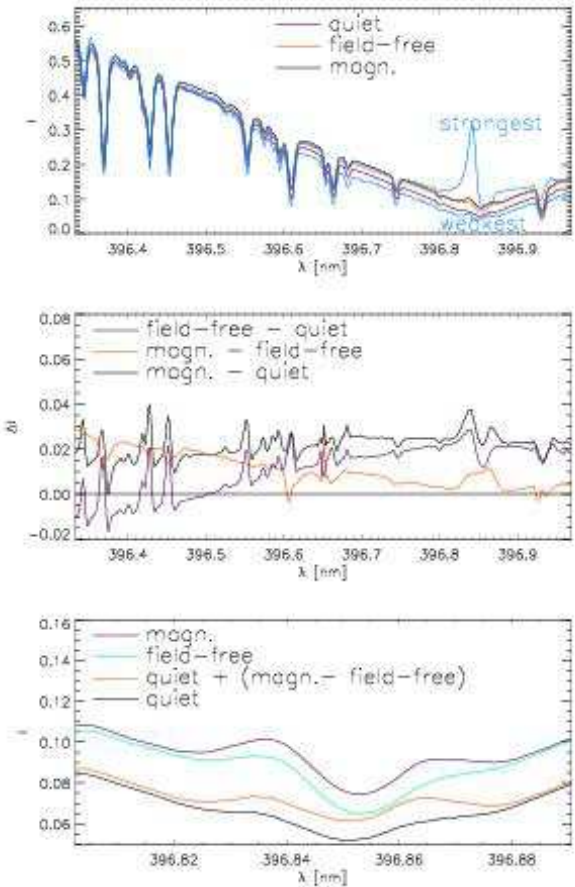}}
\caption{{\em Top}: average profiles of locations with magnetic fields, field-free emission, and quiet profiles. Strongest and weakest emission profiles are overplotted. {\em Middle}: differences between the average profiles. {\em Bottom}: close-up of the line core. The configuration of quiet profile plus the difference of magnetic and field-free profile is no observed profile.\label{fig20}}
\end{figure}

We used the value of the H-index to identify the locations in the
field of view, where at a given time some heating process must be (or have
been recently) active. We separated all profiles with high emission (H-index
$>$ 8 pm) and magnetic fields from those with emission but without
fields. As a third sample we selected all locations with strongly reduced
H-index ($<$ 7 pm). The resulting mask is shown in Fig.~\ref{fig19}. We
then averaged the profiles of each sample. The top panel of Fig.~\ref{fig20}
shows the resulting average profiles, including the profiles having
the largest, respectively, smallest H-index for comparison. To enhance the
visibility of the differences between these average profiles, we subtracted
them from each other (middle panel). This plot reveals some interesting
features. The intensity at locations of photospheric magnetic fields is seen to
be higher than the quiet profile at all wavelengths by a roughly constant
amount throughout the line wing. In the Ca II H line core, the
difference shows two peaks, where the one corresponding to H$_{\rm 2V}$ is
more pronounced. In contrast to that, the intensity for locations with
field-free emission is higher than the quiet profile only close to the core,
whereas the line wing intensity is identical. In the line core,
the asymmetry between  H$_{\rm 2V}$ and H$_{\rm 2R}$ is increased, the
latter is almost invisible. The difference between field-free and magnetic
emission shows the opposite slope in the line wing: the intensities close to
the core are similar, while in the wing the field-free locations have lower
intensity. In the line core, the difference between magnetic and field-free
emission only shows a single broad peak, where the H$_{\rm 2R}$ peak is more pronounced.
\begin{figure}
\resizebox{8.8cm}{!}{\includegraphics{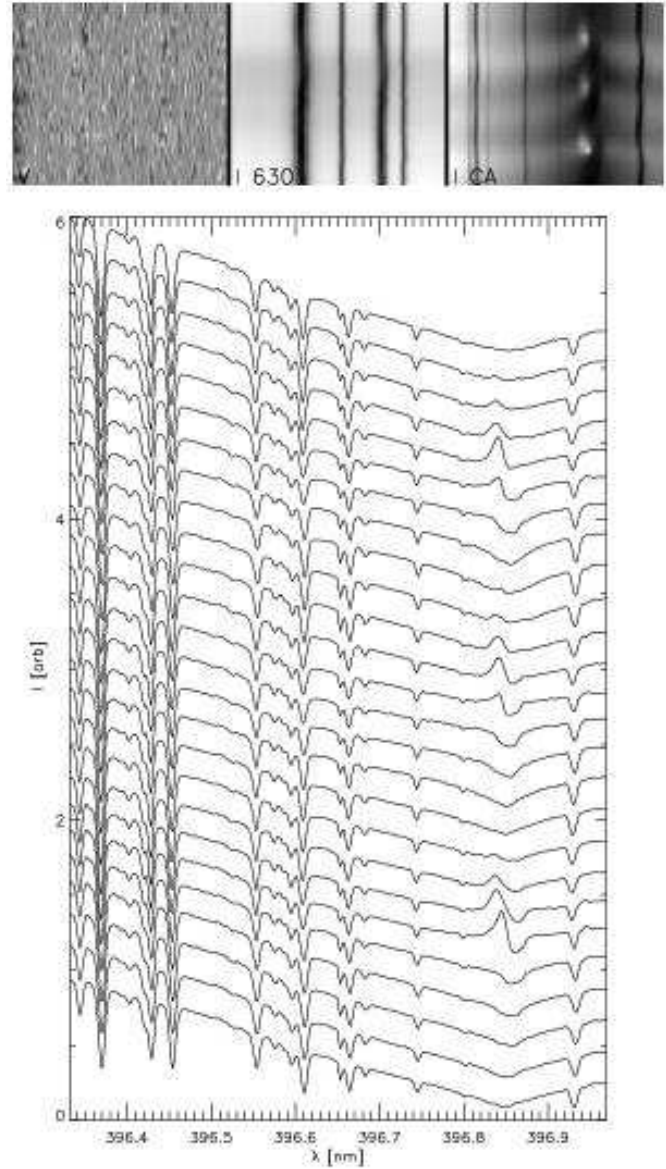}}
\caption{Temporal evolution of spectra at one fixed slit location during 500
  sec. {\em Top, left to right}: Stokes V, Stokes I 630.4\thinspace nm, Ca
  line core. The individual profiles are displayed at the bottom.\label{fig21}}
\end{figure}

If one assumes that our three samples correspond to a) a non-heated atmosphere
(quiet profiles), b) a non-magnetically heated atmosphere (field-free
emission), and, finally, c) a magnetically and non-magnetically heated
atmosphere (location of fields), one can quantify the characteristic properties of the different heating mechanisms. The non-magnetic heating shows the typical shock signature with an enhanced H$_{\rm 2V}$ peak, while leaving the wing unchanged. The magnetic heating affects the whole spectral range, raises the line wing intensity, and shows a peak symmetric to the line core. This indicates a permanent temperature rise in the upper layers more strongly than the transient emission with the shock signature in the field-free case. The increase in the line wing could be a simple consequence of the shift of the optical depth scale in the presence of the magnetic fields and not related to a heating process at all. It would be interesting to compare the profile resulting from subtracting the field-free emission from the magnetic emission with the difference between synthetic profiles assuming a non-heated chromosphere and a flux concentration embedded in the same atmosphere. \citet{solanki+etal1991} performed calculations of Ca II K spectra for several flux tube models that could be used for this purpose, if Ca II H spectra were calculated instead.

\begin{figure}
\centerline{\resizebox{8.8cm}{!}{\includegraphics{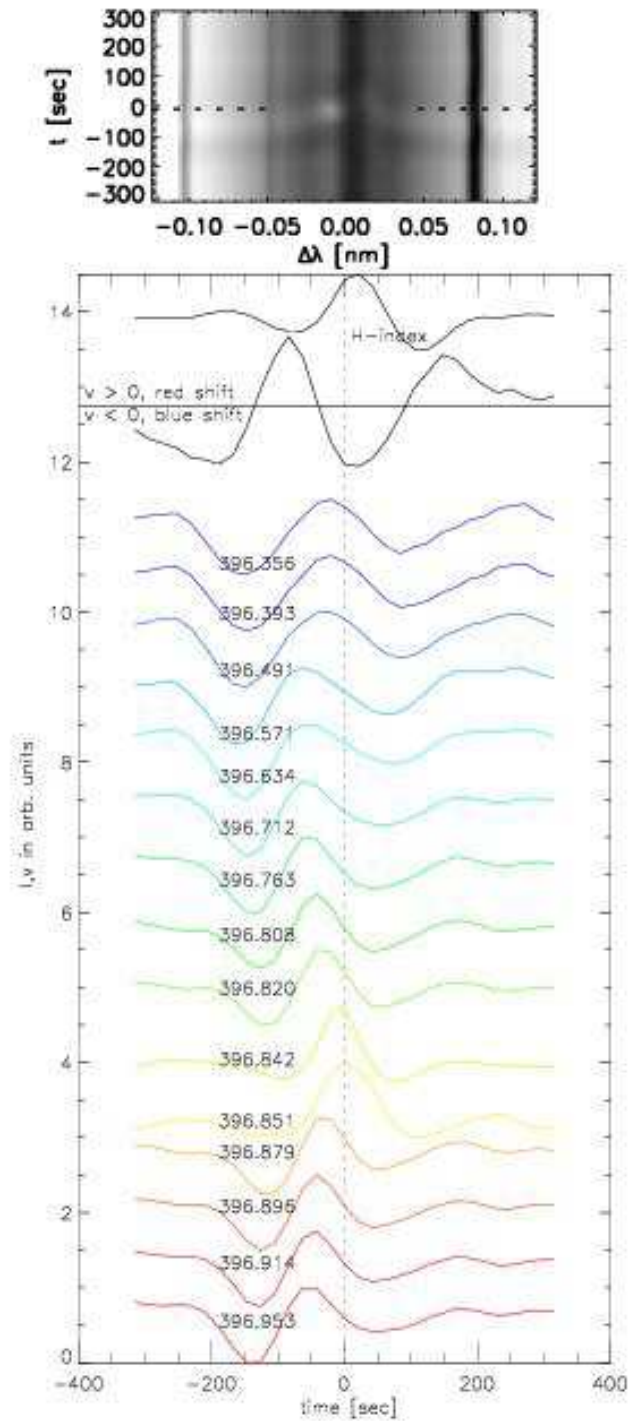}}}
\caption{Temporal evolution close to the average shock event at t=0 sec as function of time. {\em Top}: average spectra. {\em Bottom}: intensity
  in continuum bands.\label{fig22}}
\end{figure}
\section{Shock evolution\label{sect5}}
The evolution of Ca profiles during the formation and passage of shocks
has been extensively studied. The closest reproduction of the patterns observed
 was achieved by Carlsson \& Stein (1997), who employed a photospheric piston
 that generated upward propagating waves in a 1-D atmosphere. These waves
 steepened into shocks and led to the appearance of emission in H$_{\rm
   2V}$. The general pattern of the shock signature is also well known \citep[cf.][or the extensive review of Rutten \& Uitenbroek 1991]{cram+dame1983}. Figure \ref{fig21} displays one example of a
 series of three successive shock waves that are in good agreement with
 earlier descriptions for the behavior in and close to the Ca line
 core. With our large wavelength range we can also try to trace down
 the origin of the shocks leading to the strong emission in H$_{\rm 2V}$. As
 can be seen in Fig.~\ref{fig2}, both intensity increases and decreases near
 the Ca line core can be followed down to the outmost wing intensity
 observed at around 0.5\thinspace nm from the core. 

To obtain a statistically significant proof of the pattern, we determined the
locations of all profiles ($\sim$1000 cases), where the intensity of the H$_{\rm
  2V}$ peak exceeded 0.1 of I$_{\rm c}$, and averaged them. We did the same with the profiles
observed on these locations during the 300 seconds before and after the high emission. This yielded the average evolution of profiles near a shock event
(top panel of Fig.~\ref{fig22}). From these averaged profiles, we took the intensities in
continuum bands as function of time, where the ``shock'' is at t = 0 sec. The
most interesting feature is that the average shock seems to be both preceded
and followed by a reduction of intensity, where the reduction is more
pronounced before the shock event. The first intensity minimum can be seen to
travel smoothly from the wing to the core in around 100 seconds. The maximum
following it, culminating in the shock, does not seem to propagate in the same
way: in the outer wing ($\lambda < 396.571$\thinspace nm), the maximum appears
{\em later} than for, e.g., 396.571\thinspace nm. The event classified as the
shock also seems on average not to be isolated, but rather to be one of a
series of shocks. At around t=-250 sec, an initial intensity increase of
H$_{\rm 2V}$ is visible, albeit much weaker than the required intensity of 0.1.
\section{Summary \& discussion\label{sect6}}
The chromosphere as a dynamic and transient structure is hard to deal with. Whether it can be described by temporally or spatially averaged quantities or models is a matter of debate \citep[e.g.][]{kalkofen+etal1999,rammacher+cuntz2005}. Commonly its definition and the proof of its existence are taken to be the temperature rise to values above the photospheric temperature. We use this most basic definition in the following, and use the emission in the Ca II H line as indicator of the temperature rise, and thus the result of some kind of heating process responsible for it. Then two main topics have to be discussed: What are the properties and the evolution of the emission ? To what photospheric structures or events is the emission related ?

\paragraph{Properties of field-free emission} Outside strong photospheric fields, the largest emission in the Ca II H line
core appears in the shape of transient brightenings of short temporal (60
sec) and spatial (2-3$^{\prime\prime}$) extension (``bright grains''). Even if
the grains are rather short-lived, the emission takes time to build up to a
maximum and relaxes more slowly afterwards (Fig.~\ref{fig21}). The grains can
repeat  some times on the same location with a cadence of around 200
seconds. The chromospheric pattern is markedly different from the evolution of
the photospheric granules. The brightenings are mainly due to an increase of
intensity of the blue emission peak, H$_{\rm 2V}$ (cf.~Figs.~\ref{fig2} or
\ref{fig21}). This pattern is well known from the earlier spectroscopic
observations of Ca II H or K \citep[e.g.][and the references therein]{cram+dame1983,rutten+uitenbroek1991} and has been reproduced fairly well by acoustic waves steepening into shocks by \citet{carlsson+stein1997}.

\begin{figure}
\resizebox{6cm}{!}{\includegraphics{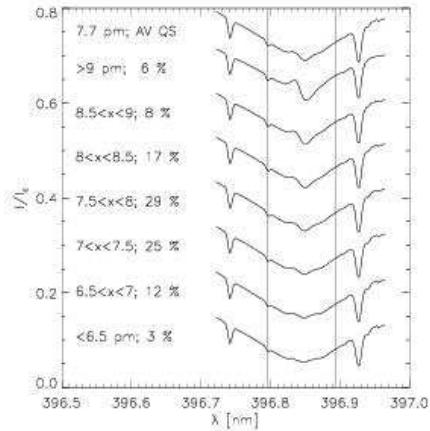}}
\caption{Average profiles corresponding to different levels of the H-index, with their relative occurence in a field-free region. The two vertical lines denote the emulated 1-{\AA} filter that yields the H-index.\label{fig24}}
\end{figure}

\paragraph{Temporal coverage of emission} \citet[][S96]{steffens+etal1996}
estimated that the chromosphere spends only 9 \% of the time in a state
leading to bright grains. This small amount led \citet{kalkofen+etal1999} to
claim that a reproduction of bright H$_{\rm 2V}$ or  K$_{\rm 2V}$ grains does
not cover the energy contained in the chromosphere, but rather only a tenth of
it. However, we remark that the selection of bright grains in S96 strongly
depended on a strict intensity threshold. To quantify the amount of time spent
in (strong) emission without imposing a threshold, we took the profiles of the
quiet Sun region in the middle of the observed FOV (cf.~Table \ref{tab2}, 2nd
row), and classified them according to increasing H-index. The average
profiles in seven bins in the H-index and their spatio-temporal area fractions
are displayed in Fig.~\ref{fig24}. For an H-index below around 7.5 pm, the
average profiles show only weak emission features. If the H-index exceeds 7.5
pm, a pronounced asymmetry of H$_{\rm 2V}$ and H$_{\rm 2R}$ is seen. Profiles
with a strongly enhanced  H$_{\rm 2V}$ peak cover around 6\% of the area, in
agreement with S96. If however the area fractions of all profiles with
emission signatures (H-index $>$7.5 pm) are added up, the ratio of profiles in emission to those
without is 60:40. Taking the full FOV, the area fraction of profiles with an
H-index above 7.5 pm is 64 \%. Another estimate can be made from Fig.~\ref{fig21}: a shock event affects usually three to four profiles, i.e.~it leads to emission for 60 to 80 seconds afterwards. If the next shock happens 180 seconds after the first one, the fraction of time spent in emission is around 70/180$\sim$40 \%. This definition of the emission from the appearance of any H$_{\rm 2V}$ peak then suggests in all estimates that the chromosphere, or more precisely, the core of the Ca II H line spends around half of the time in emission instead of 10 \%. If this emission can by modeled by a static temperature rise, or reflects a temperature rise at all, is another question. 
\paragraph{Emission in relation to magnetic fields} On locations with detected photospheric fields, a quasi-permanent increase of intensity in both emission peaks is present, in addition to similar repetitive bright grains as happen outside fields. Near to, but still outside strong photospheric magnetic fields, the emission is generally increased (cf.~Figs.~\ref{fig5} or
\ref{fig19}). Interestingly, the maximum H-index observed in the time series
is located outside of magnetic fields, which could fit with the suggestion of
\citet{kalkofen1996} that collisions between flux concentrations and granules
are responsible for the creation of bright grains. We note however that in our case it would be the interaction of a strong uni-polar network element with granulation instead of the weaker mixed-polarity fields suggested by \citet{kalkofen1996}. 

The areas with least chromospheric emission in our time series are located furthest away ($>10^{\prime\prime}$) from any magnetic fields, in the middle of the observed field of view. The spatial distribution of emission in our 1-D slit observations would comply very well with a cut through the FOV observed by \citet[][V07]{vecchio+etal2007}, if the slit would be placed across one of the field concentrations visible in their Fig.~2. The halo of enhanced emission close to the fields found in the present paper would correspond to one high-emission fibril seen in the Ca II 854.2\thinspace nm line by V07. These fibrils are interpreted to reflect the chromospheric magnetic field topology by V07, and end after around $10^{\prime\prime}$ in low-emission dark regions. 
\begin{figure}
\resizebox{8cm}{!}{\includegraphics{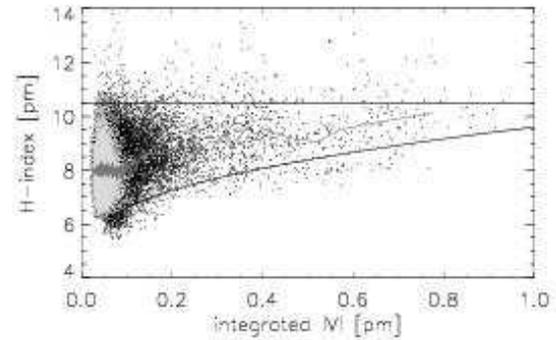}}
\caption{Scatter plot of integrated unsigned Stokes $V$ signal vs the
  H-index. {\em Black dots}: full FOV. {\em Light grey dots}: quiet area
  without network. The {\em grey line} gives the average value of H-index
  as function of $V$. Black lines outline upper and lower limits of the H-index.\label{fig23}}
\end{figure}

Like \citet{lites+etal1999}, we do not see an one-to-one correlation of
emission in calcium and photospheric fields or Stokes $V$ signal as claimed by
\citet{sivaraman+etal2000}, in none of the Figs.~\ref{fig2}, \ref{fig5}, or
\ref{fig19}. \citet{lites+etal1999} employed data very similar to ours,
spectro-polarimetry in 630\thinspace nm and spectroscopy in Ca II
H. There are several occurrences of H$_{\rm 2V}$ brightenings on locations
without any polarization signal above our detection limit of 0.15 \% of the
continuum intensity. The relation in the other direction is however rather tight: if photospheric fields are present,
the emission is enhanced and often also affects the H$_{\rm 2R}$ peak as
well (Figs.~\ref{fig2} and \ref{fig5}). To quantify the visual impression, we use the scatter plot of integrated
unsigned Stokes $V$ signal vs the H-index
(Fig.~\ref{fig23}). The scatter plot of the full FOV shows the usual
behavior
\citep[e.g.][]{skumanich+etal1975,schrijver1987,schrijver+etal1989,reza+etal2007},
a general increase of chromospheric emission with polarization signal, i.e.,
with total magnetic flux. To substantiate the claim that emission can occur
without fields, we overplotted the values of the quiet region in the middle of
the FOV (cf.Table \ref{tab2}, 2nd row) separately in light grey. It can be
clearly seen that the emission in this part of the FOV covers the {\em same
  range in the H-index} as the full FOV, from 6 to around 11 pm, but shows
only weak polarization signals. We emphasize also again the conclusion of
\citet{reza+etal2007} that the presence of magnetic flux influences the
minimum H-index, but that the maximum emission value\footnote{Excluding flares
or similar events.} seems to be independent of the magnetic flux. This gives another indirect argument that photospheric fields increase the chromospheric emission, but do not actually deliver the main contribution to it.

For the strongest concentration of magnetic flux in the field of view
observed, a stable long-lasting ($>1$ hr) network element, we 
find a  displacement of 1$^{\prime\prime}$ between magnetic flux and
highest emission, and less pronounced photospheric power. The displacement is only in one direction along the slit, which we ascribe to the field topology in the FOV. The strongest field concentration could be connected to one of opposite polarity nearby in the direction of the displacement. 
\paragraph{Properties of intensity oscillations}
The chromospheric intensity oscillations show power at all frequencies from 0 to about 10 mHz. We do not find a pronounced peak of power at 3 minutes, but a broad distribution over several frequencies. However, to address the question of heating, the average power spectrum alone is of less interest than the power spectrum of locations with strong chromospheric emission. Comparing the spatially
resolved chromospheric intensity with chromospheric and photospheric
oscillation power, or with the locations of photospheric fields, it can be
seen that strong emission in the chromosphere is always related to one of two
things (or both): magnetic fields, or high power in the photospheric velocity
oscillations. These photospheric oscillations are due to isolated small-scale
power sources in the frequency range up to the acoustic cutoff frequency of
around 5 mHz (cf.~Fig.~\ref{fig9}, lowermost panel). This agrees with the
finding of \citet{kamio+kurokawa2006} that the large-scale photospheric 3
mHz oscillations are less important for the generation of H$_{\rm 2V}$ bright grains than localized 5 mHz oscillations \citep[see also][]{hoekzema+etal2002}. The result would also be in agreement with both an impulsive excitation of waves, or a stochastic generation by the (random) superposition of large-scale wave patterns, which again would interfere positively only on some locations.

The analysis of the phase differences between the oscillations of the H$_{\rm
  2V}$ peak and the intensities at other wavelengths gives evidence that the
acting agent between photosphere and chromosphere are propagating waves with
frequencies above 2 mHz. Below 2 mHz, constant phase shifts are found that
however depend on the wavelength difference to the H$_{\rm 2V}$ peak. For
frequencies above 2 mHz, the phase differences to H$_{\rm 2V}$ allow to
determine the slope of the phase difference as function of oscillation frequency,
$\delta(\Delta\Phi)/\delta\nu$, for all wavelengths in the observed spectral
range around the Ca line core (-0.5\thinspace nm,+0.1\thinspace nm). This
quantity can be used to derive an estimate of the formation height, as the
phase difference is in the basic approximation directly proportional to a
height difference. 

\citet{woeger+etal2006} used a threshold of 150 sec in the decorrelation time to identify network areas in their narrow-band Ca II K filtergram observations. For their remaining internetwork sample, they obtained a typical time scale of around 50 sec. Figure \ref{fig16a} then suggests that the separation between network and internetwork by a decorrelation time of 150 sec should only work properly, when information from the lower chromosphere is included in the observations. Above 400 km height, the locations of magnetic fields show decorrelation times around 100 sec, whereas on the field-free locations we also find around 50 sec. We suggest using a lower threshold of around 75 sec, as at least the results agree that field-free locations should show decorrelation times below that. Figure \ref{fig16a} also implies that the pattern of inverse granulation, presumably originating at heights between 200 km and 600 km \citep[e.g.][]{lennarts+wedemeyer2005}, has a smaller decorrelation time than the photospheric granulation itself. The inverse granulation pattern also does not show up prominently in the intensity maps at different wavelengths (Fig.~\ref{fig5}).
\paragraph{Evidences of propagating waves}
That the waves responsible for the chromospheric emission are traveling from
the photosphere upwards, is shown by an analysis of the temporal evolution of
profiles at fixed locations. The brightest grains with strong H$_{\rm 2V}$
emission can all be traced back to intensity variations in the outermost line
wing observed. The pattern travels from the wing (396.35\thinspace nm) towards
the line core (396.85\thinspace nm) in around 50-100 seconds, corresponding to phase speeds between 7 and 14 kms$^{-1}$ with the height scale derived from the phase differences. Interestingly, both intensity decreases and increases can be seen traveling through the spectrum, where the brightest grains are on average preceded {\it and} followed by an intensity decrease (cf.~Fig.~\ref{fig22}). \citet{cadavid+etal2003} found a similar phenomenon, where however darkenings of G-band were either preceding {\em or} following a brightening in their rather broad-band (3{\AA}) Ca II K filtergrams. In our case, the darkening preceding a shock could appear, because most bright grains are part of a train of successive brightenings with darkenings in between.
\paragraph{Selection sensitivity}
We caution that our results may be biased by selection
effects. Our spatio-temporal field of view covers 60$^{\prime\prime} \times$ 1
hr, with a specific configuration of strong photospheric fields inside of
it. Several quantities (power spectra, average profiles, wavelength
correlation matrices) were found to be sensitive to the locations chosen
in their derivation. Even if we tried to select the locations by the commonly
used criteria of network fields and internetwork areas devoid of strong
fields, we cannot exclude the possibility that we have observed an
``atypical'' network field, because a single field concentration more or less dominated the signal for the ``magnetic'' locations. Analysis of more data will be needed to exclude such effects. Fortunately, several other time series were taken with POLIS in 2006 in observation campaigns previous to the one used here, albeit with a worse temporal sampling.
\section{Conclusions\label{sect7}}
We have analyzed a time series of intensity spectra in the chromospheric
Ca II H line and Stokes vector polarimetry in the two Fe I lines
at 630\thinspace nm. We derived the statistical properties of the emission
pattern visible in the  H$_{\rm 2V}$ and H$_{\rm 2R}$ peaks near the Ca
line core. We find that the emission is mainly due to two sources: isolated
small-scale sources of strong photospheric oscillations, or magnetic
fields. The presence of strong photospheric fields  adds only
some quasi-static emission to a pattern of transient brightenings as in
field-free locations. The emission is generally enhanced near the photospheric
fields, and smallest when furthest away from the field. The main driver of the
chromospheric emission of Ca II H are seen to be acoustic waves, propagating
upwards from the photosphere and steepening into shocks, and not the magnetic fields. We estimate that the temporal fraction of Ca II H profiles in emission is around 50\%, whereas bright H$_{\rm 2V}$ grains happen around 6\% of the time only. The emission seems to be always related to shock events, either in their increasing or decreasing phase.

We analyzed a spectral range from around 396.33\thinspace nm to
396.97\thinspace nm on the signature of the chromospheric heating
process. We suggest that these wavelengths in the line wing of Ca II H,
and the spectral lines located there, may be more helpful for the study of the
chromosphere than thought of before \citep[see also the review of][]{rutten2007}. The propagating waves leave clear traces
in the phase differences between core and wing, the wavelength correlation
matrices, or simply the wing intensity. Even an inversion of spectra assuming
LTE is able to follow a (shock) wave through the wing
(cf.~Appendix \ref{appb}). 

We find that the chromospheric heating as seen in the emission near
the Ca II H line core is dominated by propagating acoustic
waves coming from the photosphere, in agreement with the ``piston model''
of \citet{carlsson+stein1997}. Magnetic fields influence the energy
deposit by waves in and near them, which could be related to the ``magnetic
portal'' effect \citep{jefferies+etal2006}, but they seem to be of minor
importance and do not supply the energy of the chromospheric heating \citep[cf.][]{lites+etal1999,reza+etal2007}. If the emission in
Ca II H represents the chromospheric temperature increase, and thus,
the heating process, its origin are acoustic waves.

\begin{figure}
\centerline{\resizebox{8cm}{!}{\includegraphics{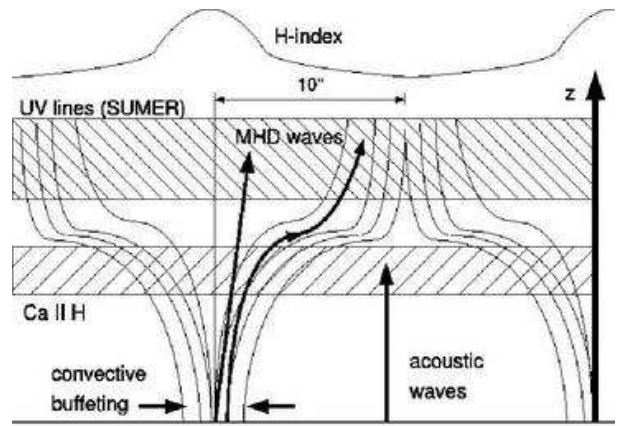}}}
\caption{Sketch of the formation height of Ca II H and the emission lines of \citet{carlsson+etal1997} in relation to a possible magnetic field topology. \label{lastfig}}
\end{figure}

The still transient emission in Ca II H -- even if we estimate around 50\% temporal filling fraction -- seems to be in contradiction to the permanent emission required to explain the SUMER observations of \citet[][CS97]{carlsson+etal1997} in far-UV spectral lines. We suggest a difference in formation height for an explanation of the discrepancy as sketched in Fig.~\ref{lastfig}. We see a changed behavior of the emission in and close to photospheric magnetic fields, in comparison to locations far from the magnetic fields ($\sim$10$^{\prime\prime}$) . This implies that in the formation height of Ca II H the magnetic fields have not yet expanded enough to fill {\it the whole volume} of the chromosphere. If the emission lines of CS97 originate from higher layers, which are magnetic everywhere and can be affected by magneto-hydrodynamic waves inside the fields, the part-time emission at only some locations of the lower chromospheric layers  ($\equiv$ acoustic heating) can be reconciled with a permanent emission in the whole volume above it ($\equiv$ magneto-acoustic heating): the magnetic field lines connect the whole volume to photospheric oscillation sources, whereas in the lower field-free chromosphere the waves can affect only a small volume on their way. The question if and where the magnetic fields form a closed canopy would then be crucial for investigations of the chromospheric heating process.

\begin{acknowledgements}
The VTT is operated by the Kiepenheuer-Institut f\"ur Sonnenphysik (KIS) at the
Spanish Observatorio del Teide of the Instituto de Astrof\'{\i}sica de Canarias (IAC). R.R.~and W.R.~acknowledge support by the Deutsche Forschungsgemeinschaft under grants SCHM 1168/8-1 and SCHM 1168/6-1, respectively. The POLIS instrument has been a joint development of the High Altitude Observatory (Boulder, USA) and the KIS. Discussions with M.~Collados are gratefully acknowledged as well.
\end{acknowledgements}

\bibliographystyle{aa}
\bibliography{references_luis_mod}

\begin{appendix}
\section{Differential refraction effects for POLIS\label{appA}}
As discussed in \citet{reardon2006}, the differential refraction in the Earths' atmosphere leads to a wavelength dependent spatial displacement. The relative displacement between two wavelengths, ($\lambda_1, \lambda_2$), can be calculated directly from the refraction index, $n(\lambda_1, \lambda_2$). The refractive index of air has been derived by various groups \citep[e.g.][and references therein]{filippenko1982,stone1996,livengood+etal1999}. For all following calculations, we have used the equations given by \citet{filippenko1982}, which yield the refractive index, $n_{\rm final} (\lambda, P,T,H)$, as function of temperature, humidity, pressure, and wavelength. The displacement in arcseconds between two wavelengths $(\lambda_1, \lambda_2)$ is then given by:
\begin{equation}
\Delta R (\lambda_1, \lambda_2) = \kappa \cdot \tan\, Z \cdot
\left[n_{\rm final}(\lambda_1) - n_{\rm final}(\lambda_2) \right],
\label{eq_diff_refrac}
\end{equation}
where $Z$ is the zenith distance of the object under examination, and $\kappa$
is the conversion coefficient from radian to arcsecsonds, $\kappa = 180/\pi
\cdot 3600 =206265$. 
\begin{figure}
\centerline{\resizebox{8.cm}{!}{\includegraphics{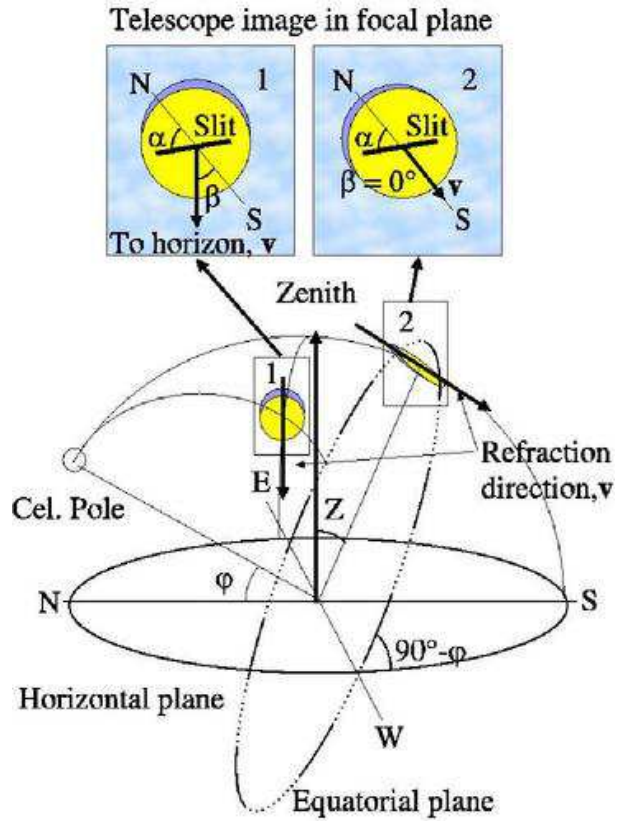}}}
\caption{Effects of differential refraction in the Earth's atmosphere:
$\phi$ is the geographical latitude, $Z$ the zenith distance of the
Sun. The differential refraction acts along the line connecting Sun center and horizon, {\bf v}, causing a vertical displacement of the solar image for different wavelengths. To derive the corresponding displacement along and perpendicular to the slit in the focal plane, the angles $\beta$, between CNS and {\bf v}, and $\alpha$, between CNS and the slit, have to be known. The image in the focal plane of the telescope has been depicted for two cases: {\bf 1} Somewhen after sunrise: CNS and {\bf v} are not parallel, $\beta \ne 0$. {\bf 2} Meridian passage of the Sun: CNS and {\bf v} are parallel, $\beta = 0^\circ$. $\alpha$ is constant for a coelostat system.\label{diff_refrac}}
\end{figure}
\begin{figure*}
\sidecaption
\resizebox{11cm}{!}{\includegraphics{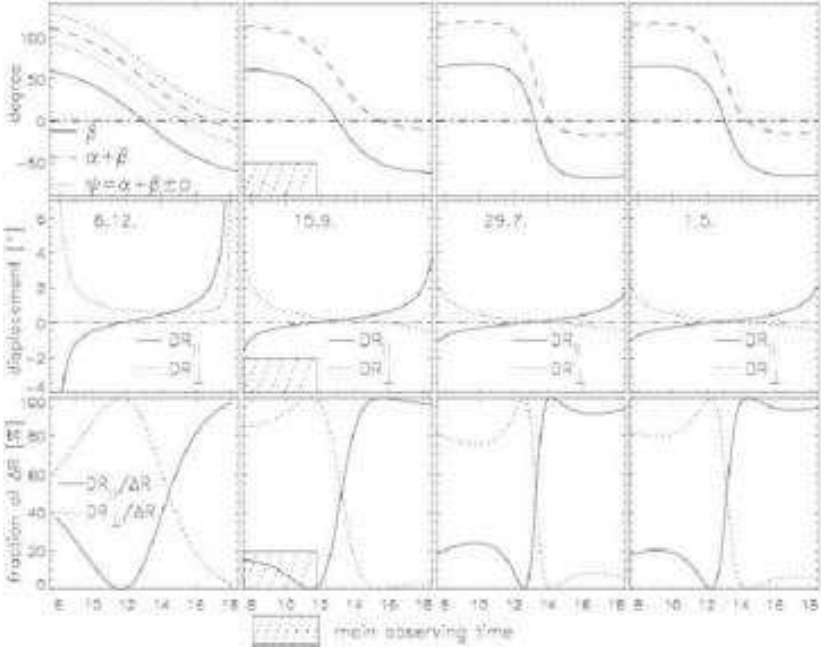}}
\caption{Angles used in the calculation, and the final displacements parallel and perpendicular to the slit for the two POLIS channels on four dates. {\em Top row}: The angle $\psi$ determines the relative fraction of $\Delta R_\perp$ and $\Delta R_\parallel$. $\beta$ denotes the orientation of celestial N-S to the local vertical; it has to pass zero at noon. The fixed slit orientation adds a constant amount $\alpha$ to $\psi$. The additional image rotation by the coelostat, $p$, can either in- or decrease $\psi$, depending on the sign of $\gamma$. {\em Middle row}: The displacements due to the differential refraction along the slit, $\rm DR_{\parallel}$ (solid), and perpendicular to the slit, $\rm DR_{\perp}$ (dotted). {\em Bottom row} The fraction of $\Delta R_{\parallel}$ and $\Delta R_{\perp}$ of the total displacement $\Delta R$.  {\em Left to right}: Parameters for 6.12., with $\gamma=\pm 90^\circ$, and for 15.9., 29.7.~and 1.5., with $\gamma = 0^\circ$. During the main observing time from 8-12 UT the displacement is mainly perpendicular to the slit.\label{angle_calc}}
\end{figure*}

However, for a dual-channel slit-spectrograph instrument with a {\em fixed}
slit like POLIS the main concern is not the absolute displacement, but the fraction perpendicular to the slit. Thus, the direction of the displacement between images in different wavelengths, $I(\lambda)$, in the focal plane has to be determined as well. The calculation of the direction of the displacement in the focal plane can be separated into three steps, a) the projection of the dispersion axis of the $I(\lambda)$ into the focal plane, b) effects due to the telescope, and c) effects due to the orientation of the slit. 
\subsection{Projection of the dispersion axis into the focal plane} 
If the position of the Sun in equatorial plane coordinates, declination, $\delta_{\odot}$, and hour angle, $t_{\odot}$ is taken from an ephemeris table, the zenith distance, $Z$, in horizontal plane coordinates can then be derived by: 
\begin{eqnarray}
Z_\odot = {\rm arccos} (\sin \phi \cdot \sin \delta_\odot + \cos \phi \cdot \cos \delta_\odot \cdot \cos t_\odot ) \; ,
\end{eqnarray}
where $\phi$ denotes the geographical latitude. With $Z$ the absolute
displacement can be calculated from Eq.(\ref{eq_diff_refrac}).

The celestial North-South (CNS) axis is defined by the tangent to the great circle through Sun center and the celestial north pole at $\delta = \delta_\odot$. The dispersion axis, {\bf v}, is analogously given by the tangent to the great circle through Sun center and the zenith point. The {\em parallactic angle}, $\beta$, then denotes the angle between the CNS axis and the dispersion axis. From the spherical triangle pole-Sun-zenith it can be derived that
\begin{equation}
\beta = {\rm arctan} \left( \frac{\sin t_\odot}{\cos \delta_\odot \cdot \tan \phi - \sin \delta_\odot \cdot \cos t_\odot }\right) \; .
\end{equation}
$\beta$ incorporates the time-dependent part of the direction of the spatial displacement, which is due to the daily solar revolution on the sky.

\subsection{Image rotation due to the telescope} A coelostat telescope system may introduce an additional image rotation in the focal plane. A displacement of the first coelostat mirror by the angle $\gamma$ from the {\em terrestrial} N-S axis leads to a constant image rotation, $p$, which is given by:
\begin{equation}
p = -{\rm arcsin} \left( \frac{\cos \phi \cdot \sin \gamma}{\cos
\delta_{\odot}} \right) + \gamma \;.
\end{equation}

\subsection{Instrument orientation in the focal plane}
Finally, for a slit-spectrograph the orientation of the slit relative to CNS has to be considered. This will be denoted by the angle $\alpha$ between the slit and the CNS axis as defined above. For a coelostat system, $\alpha$ and $p$ are constant during the day and depend only on the telescope and instrument geometry. The angle, $\psi$, between the slit and the dispersion axis is then given by the addition of all contributions:
\begin{equation}
\psi = \beta + (p + \alpha) \; ,
\end{equation}
where the parentheses indicate the contribution due to telescope and instrument orientation. The displacements perpendicular, $\Delta R_\perp$, and parallel to the slit, $\Delta R_\parallel$, are then given by:

\begin{eqnarray}
\begin{pmatrix} \Delta R_{\perp} \cr \Delta R_{\parallel} \cr \end{pmatrix} =
\begin{pmatrix} \sin \psi  \cr \cos \psi \end{pmatrix}\cdot \Delta
R(\lambda_1, \lambda_2), \label{dr_final}
\end{eqnarray}
with $\Delta R(\lambda_1, \lambda_2)$ from Eq.~(\ref{eq_diff_refrac}).

Figure \ref{angle_calc} shows the resulting displacements for the two POLIS channels at 396\thinspace nm and 630\thinspace nm on four dates. For observations on 24th of July, at around UT 8:00, a displacement perpendicular to the slit of around 2$^{\prime\prime}$ can be read off.
\section{LTE inversion examples\label{appb}}
\begin{figure}[b]
\centerline{\resizebox{6cm}{!}{\includegraphics{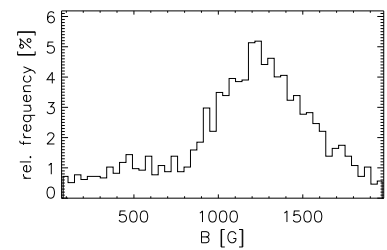}}}
\caption{Histogram of field strength in the inversion.\label{fig6}}
\end{figure}
Two inversions were performed, one in the full field-of-view (FOV) using only
the spectra of the red channel, the second in a restricted area including the
Ca intensity spectra in the fit as well. The FOV contained only few
locations with magnetic fields, which mainly were stable network elements
persisting throughout the 1-h time series. The distribution of field strengths
in Fig.~\ref{fig6} thus shows mainly fields around 1.3 kG. Note that due to
the fixed slit position the same flux concentration contributes multiple
 (up to 150 !) times.
\begin{figure*}
\centerline{\resizebox{17cm}{!}{\includegraphics{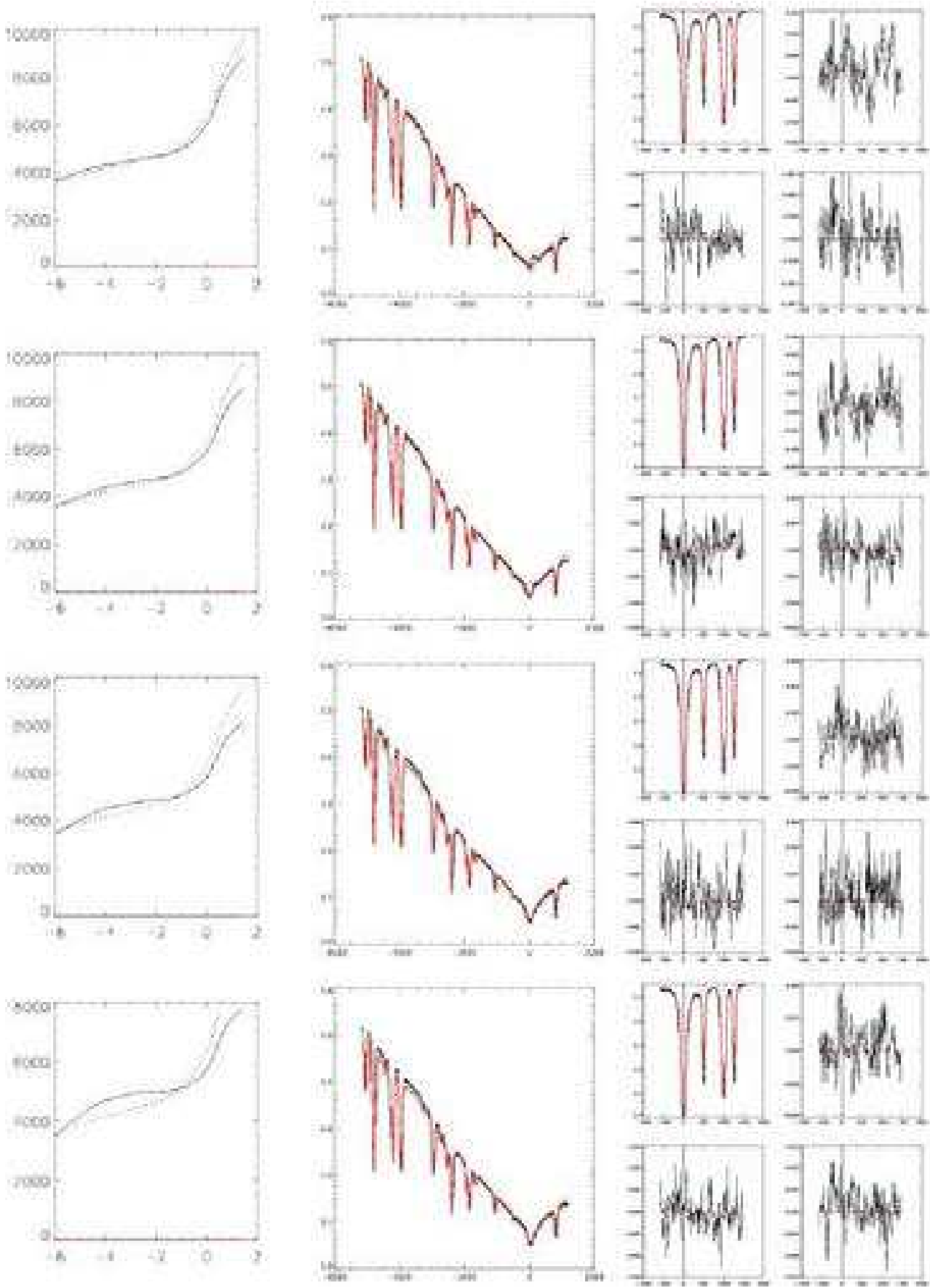}}}
\caption{Examples of the LTE inversion. {\em Left}: temperature stratification.  {\em Middle}: Ca spectra ({\em black}), best-fit profiles ({\em red}).  {\em Right}: same for 630\thinspace nm,   {\em clockwise}: IQUV.\label{lte1}}
\end{figure*}

\begin{figure*}
\centerline{\resizebox{17cm}{!}{\includegraphics{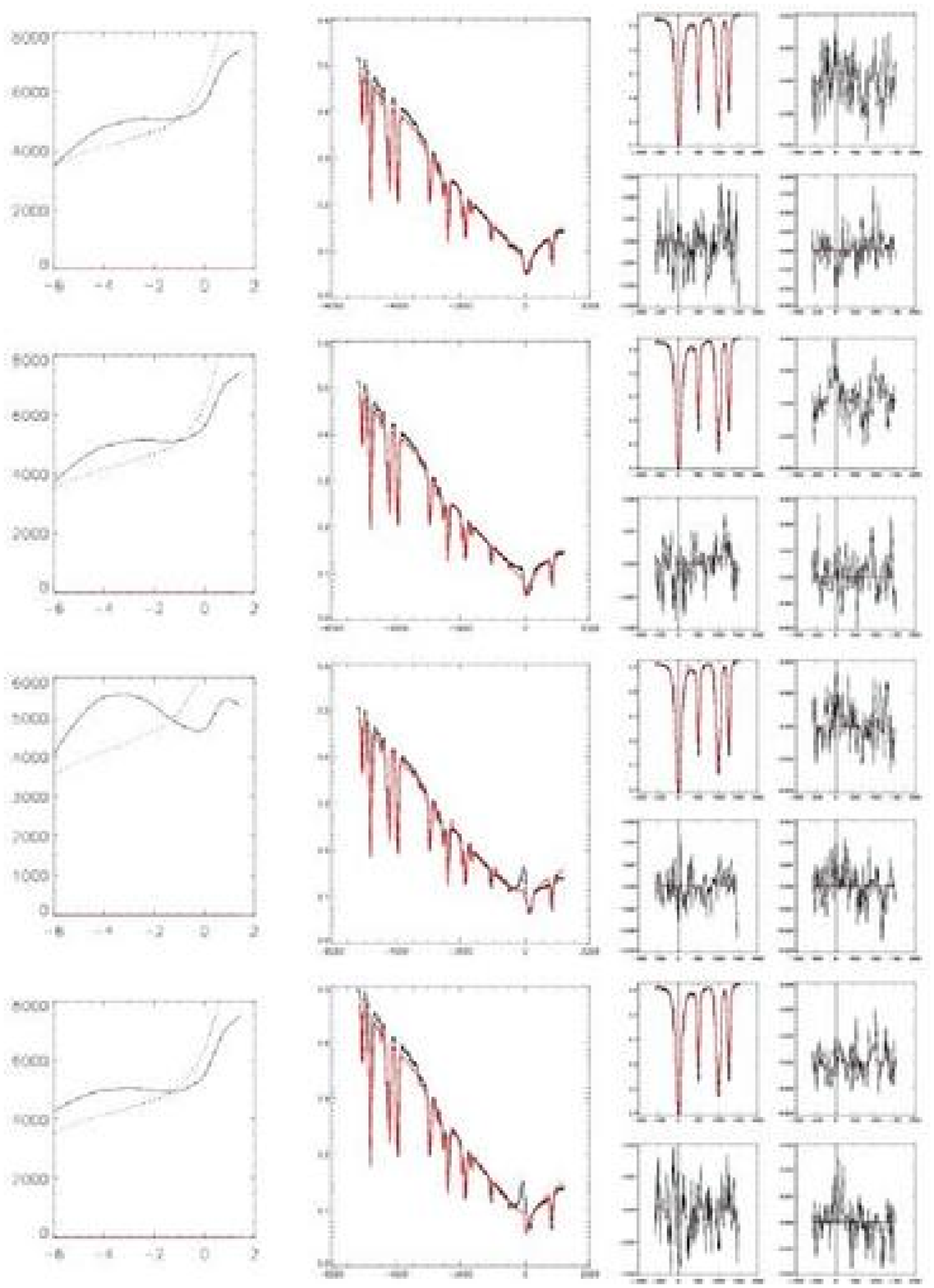}}}
\caption{Same as previous Fig.\label{lte2}}
\end{figure*}

Figures \ref{lte1} and \ref{lte2} show examples of the LTE inversion results
with the SIR code, where the complete Ca II H intensity spectrum and
Stokes $IQUV$ of the 630\thinspace nm channel were used in the fit. The
inversion scheme employed a straylight contamination with the average quiet
Sun profile, and a field-free inversion component. The straylight contribution
was generally always larger than 80 \%. As the amount of chromospheric heating
contained in the straylight profile is non-zero (two reversals in Ca II
H intensity profile, Fig.~\ref{fig24}), the results of the field-free inversion component will tend to underestimate the temperature increase.

The profiles were taken from the same spatial location and show the temporal
evolution during 168 seconds. The NLTE effects in the Ca II H line core
can of course not be recovered. However, the inversion still yields a
temperature increase in the upper atmospheric layers that travels upwards in
optical depth with time. The amplitude and properties of this temperature
increase are actually not governed by the line core, but the line wing
intensities close to the Ca II H core, where LTE maybe still
applies. Note that on this display scale the actually still existing mismatch
between fit and observations in the 630\thinspace nm channel can not be seen
at all. To summarize, it seems feasible  very well to reproduce the
photospheric spectra and the Ca line wing at the same time with a
``reasonable'' temperature stratification including a chromospheric temperature rise.
\end{appendix}
\end{document}